\documentclass[onecolumn]{aastex631}

\shorttitle{AF~Lep~b VLTI/GRAVITY}
\shortauthors{Balmer et al.}
\graphicspath{{./}{figures/}}
\definecolor{mylinkcolor}{RGB}{16, 37, 110} 
\hypersetup{linkcolor=mylinkcolor,citecolor=mylinkcolor,urlcolor=mylinkcolor}

\newenvironment{rotatepage}%
    {\clearpage\pagebreak[4]\global\pdfpageattr\expandafter{\the\pdfpageattr/Rotate 90}}%
    {\clearpage\pagebreak[4]\global\pdfpageattr\expandafter{\the\pdfpageattr/Rotate 0}}%

\accepted{Nov. 8th to the Astronomical Journal}

\begin{document}

\title{VLTI/GRAVITY Observations of AF~Lep~b: \\Preference for Circular Orbits, Cloudy Atmospheres, and a Moderately Enhanced Metallicity}

\author[0000-0001-6396-8439]{William O. Balmer}
\correspondingauthor{William O. Balmer}
\email{wbalmer1@jhu.edu}
\altaffiliation{Johns Hopkins University George Owen Fellow}
\affiliation{Department of Physics \& Astronomy, Johns Hopkins University, 3400 N. Charles Street, Baltimore, MD 21218, USA}
\affiliation{Space Telescope Science Institute, 3700 San Martin Drive, Baltimore, MD 21218, USA}

\author[0000-0003-4557-414X]{Kyle Franson}
\altaffiliation{NSF Graduate Research Fellow}
\affiliation{Department of Astronomy, The University of Texas at Austin, Austin, TX 78712, USA}

\author[0000-0001-5250-9384]{Antoine Chomez}
\affiliation{LESIA, Observatoire de Paris, PSL, CNRS, Sorbonne Universit\'e, Universit\'e de Paris, 5 place Janssen, 92195 Meudon, France}
\affiliation{Univ. Grenoble Alpes, CNRS, IPAG, 38000 Grenoble, France}

\author[0000-0003-3818-408X]{Laurent Pueyo}
\affiliation{Space Telescope Science Institute, 3700 San Martin Drive, Baltimore, MD 21218, USA}

\author[0000-0002-5823-3072]{Tomas Stolker}
\affiliation{Leiden Observatory, Leiden University, Einsteinweg 55, 2333 CC Leiden, The Netherlands}

\author[0000-0002-6948-0263]{Sylvestre Lacour}
\affiliation{LESIA, Observatoire de Paris, PSL, CNRS, Sorbonne Universit\'e, Universit\'e de Paris, 5 place Janssen, 92195 Meudon, France}

\author{Mathias Nowak}
\affiliation{Institute of Astronomy, University of Cambridge, Madingley Road, Cambridge CB3 0HA, United Kingdom 939}
\affiliation{Kavli Institute for Cosmology, University of Cambridge, Madingley Road, Cambridge CB3 0HA, United-Kingdom}

\author[0000-0002-9792-3121]{Evert Nasedkin}
\affiliation{Max Planck Institute for Astronomy, K\"onigstuhl 17, 69117 Heidelberg, Germany} 
\affiliation{School of Physics, Trinity College Dublin, The University of Dublin, Dublin 2, Ireland}

\author[0000-0003-2202-1745]{Markus J. Bonse}
\affiliation{ETH Zurich, Institute for Particle Physics \& Astrophysics, Wolfgang-Pauli-Str. 27, 8093 Zurich, Switzerland}
\affiliation{Max Planck Institute for Intelligent Systems, Max-Planck-Ring 4, 72076 Tubingen, Germany}

\author[0000-0002-5113-8558]{Daniel Thorngren}
\affiliation{Department of Physics \& Astronomy, Johns Hopkins University, 3400 N. Charles Street, Baltimore, MD 21218, USA}

\author[0000-0002-6217-6867]{Paulina Palma-Bifani}
\affiliation{Laboratoire Lagrange, Universite\' Cote d'Azur, CNRS, Observatoire de la Cote d'Azur, 06304 Nice, France}
\affiliation{LESIA, Observatoire de Paris, PSL, CNRS, Sorbonne Universit\'e, Universit\'e de Paris, 5 place Janssen, 92195 Meudon, France}

\author[0000-0003-4096-7067]{Paul Molli\`ere}
\affiliation{Max Planck Institute for Astronomy, K\"onigstuhl 17, 69117 Heidelberg, Germany}

\author[0000-0003-0774-6502]{Jason J. Wang}
\affiliation{Center for Interdisciplinary Exploration and Research in Astrophysics (CIERA) and Department of Physics and Astronomy, Northwestern University, Evanston, IL 60208, USA}

\author[0000-0002-3726-4881]{Zhoujian Zhang} 
\affiliation{Department of Astronomy \& Astrophysics, University of California, Santa Cruz, CA 95064, USA}
\altaffiliation{NASA Sagan Fellow}

\author[0000-0003-2670-9387]{Amanda Chavez}
\affiliation{Center for Interdisciplinary Exploration and Research in Astrophysics (CIERA) and Department of Physics and Astronomy, Northwestern University, Evanston, IL 60208, USA}

\author[0000-0003-2769-0438]{Jens Kammerer}
\affiliation{European Southern Observatory, Karl-Schwarzschild-Str. 2, 85748 Garching, Germany}

\author[0000-0002-3199-2888]{Sarah Blunt}
\affiliation{Center for Interdisciplinary Exploration and Research in Astrophysics (CIERA) and Department of Physics and Astronomy, Northwestern University, Evanston, IL 60208, USA}
\affiliation{Department of Astronomy \& Astrophysics, University of California, Santa Cruz, CA 95064, USA}

\author[0000-0003-2649-2288]{Brendan P. Bowler}
\affiliation{Department of Astronomy, The University of Texas at Austin, Austin, TX 78712, USA}

\author{Mickael Bonnefoy}
\affiliation{Univ. Grenoble Alpes, CNRS, IPAG, 38000 Grenoble, France}

\author{Wolfgang Brandner}
\affiliation{Max Planck Institute for Astronomy, K\"onigstuhl 17, 69117 Heidelberg, Germany}

\author{Benjamin Charnay}
\affiliation{LESIA, Observatoire de Paris, PSL, CNRS, Sorbonne Universit\'e, Universit\'e de Paris, 5 place Janssen, 92195 Meudon, France}

\author{Gael Chauvin}
\affiliation{Laboratoire Lagrange, Universite\' Cote d'Azur, CNRS, Observatoire de la Cote d'Azur, 06304 Nice, France}

\author{Th. Henning}
\affiliation{Max Planck Institute for Astronomy, K\"onigstuhl 17, 69117 Heidelberg, Germany}

\author{A.-M. Lagrange}
\affiliation{Univ. Grenoble Alpes, CNRS, IPAG, 38000 Grenoble, France}
\affiliation{LESIA, Observatoire de Paris, PSL, CNRS, Sorbonne Universit\'e, Universit\'e de Paris, 5 place Janssen, 92195 Meudon, France}

\author[0000-0001-9431-5756]{Nicolas Pourré}
\affiliation{Univ. Grenoble Alpes, CNRS, IPAG, 38000 Grenoble, France}

\author{Emily Rickman}
\affiliation{European Space Agency (ESA), ESA Office, Space Telescope Science Institute}

\author{Robert De Rosa}
\affiliation{European Southern Observatory, Alonso de Co\'rdova 3107, Vitacura, Santiago, Chile}

\author{Arthur Vigan}
\affiliation{Aix Marseille Univ, CNRS, CNES, LAM, Marseille, France}

\author{Thomas Winterhalder}
\affiliation{European Southern Observatory, Karl-Schwarzschild-Stra\ss e 2, 85748 Garching, Germany}

\begin{abstract}

Direct imaging observations are biased towards wide-separation, massive companions that have degenerate formation histories. Although the majority of exoplanets are expected to form via core accretion, most directly imaged exoplanets have not been convincingly demonstrated to follow this formation pathway. We obtained new interferometric observations of the directly imaged giant planet AF~Lep~b with the VLTI/GRAVITY instrument. We present three epochs of $\sim50\,\mu\mathrm{as}$ relative astrometry and the \textit{K}-band spectrum of the planet for the first time at a resolution of $R=500$. Using only these measurements, spanning less than two months, and the Hipparcos-Gaia Catalogue of Accelerations, we are able to significantly constrain the planet's orbit; this bodes well for interferometric observations of planets discovered by \textit{Gaia} DR4. Including all available measurements of the planet, we infer an effectively circular orbit ($e<0.02, 0.07, 0.13$ at $1, 2, 3\,\sigma$) in spin-orbit alignment with the host, and a measure a dynamical mass of $M_\mathrm{p}=3.75\pm0.5\,M_\mathrm{Jup}$. Models of the spectrum of the planet show that it is metal rich ([M/H]$=0.75\pm0.25$), with a C/O ratio encompassing the solar value. This ensemble of results show that the planet is consistent with core accretion formation.

\end{abstract}



\section{Introduction} \label{sec:intro}

\subsection{Tracing the formation of gas giants with direct observations}

\par Gas giants play a central role in shaping the formation and evolution of planetary systems in general, and our own Solar System in particular \citep[e.g.][]{Levison2003, Raymond2014, Childs2019, Horner2020}. Once they have formed, giant planets dominate the subsequent production and dynamical evolution of dust, planetesimals, and inner terrestrial planets, likely dictating the volatile content of terrestrial planets \citep[e.g.][]{Raymond2008, Raymond2012, Raymond2017, Sotiriadis2018, Bitsch2020, Horner2020b}. It has even been suggested recently that the presence of outer giant planets is correlated with inner terrestrial planets \citep{Rosenthal2022}. As such, it is important to understand the formation and evolution of gas giants in order to better understand the occurrence of Earth-like planets and the emergence of life in the universe. Gas giants themselves are rich laboratories: they host complex moon systems \citep{Morrison1982}, bear dramatic storm systems and weather patterns \citep{Marcus1993}, are obscured by clouds of varying species \citep{Helling2019}, and have interiors with poorly constrained structure and composition whose investigation drives advancement in physics modeling and experimentation \citep{Hubbard2002, Guillot2005}.

\par High-contrast imaging observations enabled by adaptive optics and coronagraphic instruments have resulted in the discovery of a growing number of directly imaged planets ($M<13\,M_\mathrm{Jup}$) in wide orbits ($a>10\,\mathrm{au}$) around young (${<}100\,\mathrm{Myr}$) pre-main sequence stars \citep[for a review, see][]{Currie2023}. These objects appear much rarer than shorter separation gas giants \citep[e.g.][and references therein]{Nielsen2019, Vigan2021} and their formation histories are often debated. 


Demographic studies from the larger sample of older radial velocity (RVs) gas giants have shown that, 1) the fraction of giant planets orbiting a star increases with stellar metallicity, tracing core accretion, as more solids are found in the planet forming disks of higher metallicity stars \citep{Gonzalez1997, Santos2004, Fischer2005, Mordasini2012b} and 2) planets with masses below about $4\,M_\mathrm{Jup}$ are found orbiting metal rich host stars, whereas objects with masses above $4\,M_\mathrm{Jup}$ exhibit a much broader range of host-star metallicity \citep{Santos2017, Schlaufman2018}. The bulk of gas giants appear to orbit near their host star's ice lines, $\sim$3--10\,au around solar type stars \citep{Fernandes2019, Wittenmyer2020, Fulton2021, Lagrange2023}, albeit with a large scatter ranging from ultra hot Jupiters to planets with periods $\gtrsim$1000 years. Bulk density measurements from transiting planets with RV masses have elucidated a planetary mass vs planetary metallicity trend \citep{Guillot2006, Miller2011, Thorngren2016} that is explainable by core accretion \citep[e.g.][]{Hasegawa2018}. At higher masses, planets are generally less metal rich because they have accreted a larger fraction of H/He gas \citep{Thorngren2019}, though these samples are based on short period planets that are much closer to their stars than directly detected objects. Despite remaining open questions, core accretion formation is the predominant theory of planet formation for the sample of indirectly detected planets \citep[e.g.][]{Raymond2022}.

\par Many objects in the current sample of directly imaged exoplanets are broadly consistent with formation in 1) the low probability, high mass tail of core accretion model distributions, and subsequent dynamical scattering outwards \citep[e.g.][]{Mordasini2009, Marleau2019, Emsenhuber2021a, Emsenhuber2021b}, 2) the low probability, low mass tail of molecular core fragmentation and early dynamical capture distributions \citep[e.g.][]{Chabrier2003, Padoan2004, Boyd2005, Bate2009, Bate2012}, or perhaps 3) as instances of disk instability and fragmentation \citep[e.g.][]{Boss1997, Kratter2010, Forgan2018}. There is growing consensus from dynamical studies tracing the eccentricity and obliquity distributions of these systems that directly imaged planets interior to $\sim50-100\,\mathrm{au}$ formed within a disk \citep[e.g.][]{Bowler2020, Bowler2023, Sepulveda2024}, but it is still unclear whether these objects formed via core accretion or disk fragmentation. 
\par The challenge for direct imaging is twofold, to push to observe planets that are representative of the bulk of the exoplanet population, and to disentangle the formation histories of known imaged planets. Direct observations, combined with model independent mass measurements, can then seek to expand on the observational trends to inform planet formation models with greater fidelity.

\subsection{Accessing solar system scales with interferometry}

\par In addition to classical direct imaging using coronagraphic imaging, or molecular mapping via high-resolution cross correlation, optical interferometry has begun to provide direct observations of gas giant planets with the advent of the Very Large Telescope Interferometer (VLTI) GRAVITY instrument \citep{GRAVITYCollaboration2017} and the ExoGRAVITY survey \citep{Lacour2020}. GRAVITY is a fiber fed instrument with dedicated fringe tracking. Informed by radial velocity observations and proper motion anomalies \citep{Grandjean2019, Lagrange2020}, GRAVITY has helped validate and reveal two newly directly detected planets within $10\,\mathrm{au}$, $\beta$ Pictoris c \citep[$2.7\,\mathrm{au}$;][]{Nowak2020} and HD~206893~c \citep[$3.5\,\mathrm{au}$;][]{Hinkley2023}. With absolute astrometry from \textit{Gaia}, GRAVITY has detected a handful of brown dwarf companions at very close separations \citep[$60-200\mathrm{mas}$;][]{Pourre2024, Winterhalder2024}. It has also exquisitely characterized many known directly imaged giant planets and brown dwarf companions \citep{GRAVITYCollaboration2019, GravityCollaboration2020, Kammerer2021, Lacour2021, Wang2021a, Balmer2023, Blunt2023, Balmer2024, Nasedkin2024}. 

\par GRAVITY provides the most precise relative astrometry of directly detected planets yet measured, as well as \textit{K}-band spectroscopy (in three modes, with R=50, 500, and 5000). The \textit{K}-band emission of substellar objects is shaped by absorption from H$_2$O, CO, and CH$_4$. The capability to conduct spectroscopy at close separations has enabled the precise estimation of the atmospheric abundances of a number of directly imaged planets \citep{Molliere2020}. GRAVITY's main limitation is its very small field of view ($\sim60\,\mathrm{mas}$), which necessitates precise fiber pointing (and therefore some prior knowledge of the planet's position) to ensure the planet's emission is coupled into the science fiber \citep[see][for the characterization of the contrast performance of the techniques used in this paper and previous work]{Pourre2024}.

\par The release of \textit{Gaia}~DR4 is expected to contain a list of astrometrically detected exoplanets\footnote{\url{https://www.cosmos.esa.int/web/gaia/release}}, which would provide prior knowledge of the positions of a large sample of planets that could be amenable to imaging with GRAVITY. Previous estimates of \textit{Gaia}'s exoplanet detection capabilities have indicated that as many as 21,000 long period, giant exoplanets could be found over the mission's lifetime \citep{Perryman2014}. GRAVITY is expected to uniquely detect and characterize a subset of these astrometric planet candidates in young moving groups \citep{Winterhalder2024, Pourre2024}. This would open a unique window into atmospheric characterization of giant planets near their snow lines. The question then becomes, `to what degree do we expect to understand the orbits, atmospheric composition, and fundamental properties of planets that can only be detected (in the near term) with \textit{Gaia} and GRAVITY?'

\subsection{AF~Leporis~b}

\par Proper motion anomalies between \textit{Hipparcos} and \textit{Gaia} \citep{Brandt2018, Kervella2019, Brandt2021, Kervella2022} have been leveraged to directly image new companions with a higher detection rate than blind surveys \citep[e.g.,][]{Currie2020, Chilcote2021, Bonavita2022, Franson2023a, Currie2023, Li2023}. Most notably, after uninformative surveys the decades prior had returned non-detections \citep{Biller2013, Stone2018, Nielsen2019, Launhardt2020}, proper motion anomaly informed searches have uncovered a giant planet orbiting the young star AF~Lep (HD~35850, HIP~25486, \textit{Gaia}~DR3~3009908378049913216) that has the lowest dynamically measured mass of any directly imaged exoplanet \citep{DeRosa2023, Franson2023c, Mesa2023}. The mass estimates for the planet at the time of discovery varied by about a Jupiter mass between datasets \citep[$4.3^{+2.9}_{-1.2}\,M_\mathrm{Jup}$,][]{DeRosa2023}, \citep[$3.2^{+0.7}_{-0.6}\,M_\mathrm{Jup}$,][]{Franson2023c}, \citep[$5.237^{+0.085}_{-0.10}\,M_\mathrm{Jup}$,][]{Mesa2023}, depending on the orbital coverage of the initial observations. The ensemble of discovery data was subsequently analyzed together, providing another estimate of the planet's dynamical mass \citep[$2.8^{+0.6}_{-0.5}\,M_\mathrm{Jup}$,][]{Zhang2023}. The planet was recovered in archival observations dating back to 2011 using explainable machine learning assisted starlight subtraction, which significantly extended the baseline of the planet's measured orbital motion, and constrained the orbit to low eccentricities \citep{Bonse2024}. 
\par Initial estimates for the temperature and spectral type of the planet varied anywhere between $700-1200\,\mathrm{K}$, placing the planet along the low-surface gravity L-T transition, where spectral modeling is particularly challenging \citep{Marley2015}. Nevertheless, by combining the suite of discovery spectrophotometry, two studies have indicated that the planet appears to have a temperature between $700-900\,\mathrm{K}$ and an atmosphere that is enriched in metals compared to the solar value \citep{Zhang2023, Palma-Bifani2024}. The planet was directly imaged at 4.4\,\textmu m using \textit{JWST}/NIRCam coronagraphy, and the suppression of its flux at these wavelengths provided strong evidence that its atmosphere is in chemical disequilibrium \citep{Franson2024}. It has been suggested that the luminosity and dynamical mass of the planet are consistent with a delayed formation \citep[][that is, formation a few Myr after the host star formation]{Franson2023c, Zhang2023, Zhang2024}. 
\par The sensitivity of the planet's dynamical mass estimate to the viewing geometry, as well as the relatively unconstrained methane abundance given the wavelength coverage of existing observations motivated our follow-up of this system with VLTI/GRAVITY.


\par Here we present the results of our new interferometric observations of AF~Lep~b with VLTI/GRAVITY. Relative astrometry at $\sim50\,\mu\mathrm{as}$ precision near periastron constrains the low eccentricity of the planet's orbit with a time baseline of only 2 months, and coupled with the archival recovery of the planet near apoastron, places a strong upper limit on the eccentricity. The system provides an example of the excellent orbital precision that could be expected for planets detected astrometrically by \textit{Gaia} and directly confirmed by GRAVITY. The \textit{K}-band spectrum of the planet shows prominent methane absorption features that constrain the temperature, composition, and chemical disequilibrium of the atmosphere. We discuss future improvements necessary to better characterize the planet's atmosphere and bulk properties. We argue that the current ensemble of results are consistent with expectations from core accretion formation models. We conclude by looking forward to anticipated observations of \textit{Gaia} discovered planets with VLTI/GRAVITY.

\section{Observations and Data Reduction} \label{sec:obs}
\subsection{VLTI/GRAVITY} \label{subsec:gravity}
\par We observed AF~Lep~b the night of UT 2023 November 2, UT 2023 November 24, and UT 2023 December 24 using the GRAVITY instrument \citep{GRAVITYCollaboration2017} at the European Southern Observatory (ESO) Very Large Telescope Interferometer (VLTI). We used the four 8.2~m Unit Telescopes (UTs) and the dual-field mode of GRAVITY. The fringe tracker \citep{Lacour2019, Nowak2024a}, used to maintain the stability of the interferometric observables, was placed at the location of the host star, while the science fiber was placed at the predicted location of the planet and integrated to detect its fringe pattern. Our observation on 2023-11-02 was taken in off-axis mode, and was complemented with an observation of the binary HD~25535 AB to calibrate the phase of the coherent flux \citep{Nowak2024b}, and with a single-field on-axis observation of the host star to calibrate the amplitude. Our observations on 2023-11-24 and 2023-12-24 were taken in on-axis mode, moving the science fiber periodically between the host star and the planet. The on-star observations were used to calibrate both the phase and the amplitude of the coherent flux, meaning that no observation of a separate calibrator was required. Conditions on 2023-11-02 were rather poor ($\mathrm{seeing}=1\farcs1-1\farcs9$) while conditions on 2023-11-24 and 2023-12-24 were good ($\mathrm{seeing}=0\farcs4-0\farcs8$). Our predictions for the location of the planet were based on all previously available relative  astrometry\footnote{We fit 10,000 orbits to the available relative astrometry of the planet ahead of each observation using the Orbits For The Impatient (OFTI) algorithm via the \texttt{orbitize!} package \citep{Blunt2017, Blunt2020}. We then predicted the location of the planet during a given week of VLTI UT runs based on these orbits.}, and achieved photon coupling efficiencies into the single mode fiber of $\gamma>0.98$. This coupling efficiency is an analytic estimate of the throughput of the source depending on the displacement between the source and the center of the science fiber, and a derivation is given in Appendix A of \citep[][]{Wang2021a}. Table \ref{tab:obslog} records our observing log.

\begin{figure*}
    \centering
    \includegraphics[width=\textwidth]{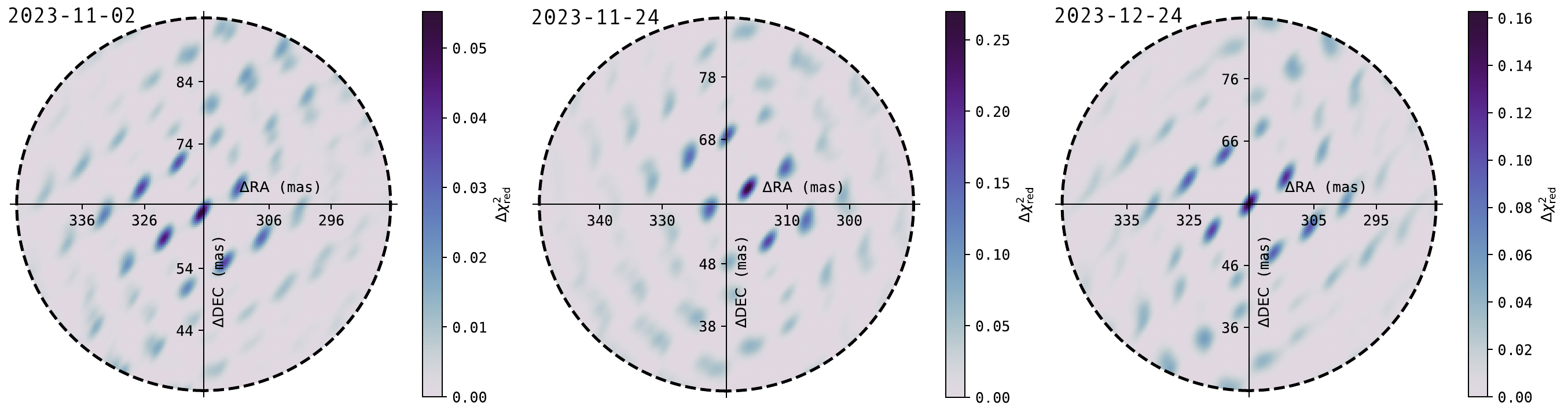}
    \caption{Detections of AF~Lep~b within the VLTI/GRAVITY fiber field of view. Each panel visualizes the $z_{\rm red}=\Delta\chi^2_{\rm red}$ calculated for various positions ($\Delta$RA and $\Delta$Dec in milliarcseconds, i.e. the displacement with respect to the measured position of the host star AF~Lep~A), fitting the interferometric observables with a polynomial model of the transmission and stellar halo contribution and a coherent point source with the spectral signature of a T-dwarf.
    Each epoch is presented chronologically, left to right: 2023-11-02, 2023-11-24, and 2023-12-24. The dashed black circle indicates the effective fiber field-of-view, $\sim60\,\mathrm{mas}$. The origin is the placement of the science fiber on-sky for a given observation, a prediction based on the previous available orbit fit. The strongest peak in the $\chi^2$ map reveals the position of the companion, with lower likelihood peaks surrounding this central peak. The shape and distribution of the likelihood depends on the \textit{u}-\textit{v} plane coverage of the observations.}
    \label{fig:detections}
\end{figure*}

\par 
The raw data were reduced using the Public Release 1.5.0 (1 July 2021\footnote{\url{https://www.eso.org/sci/software/pipelines/gravity}}) of the ESO GRAVITY pipeline \citep{Lapeyrere2014}, up to the ``astroreduced" intermediate data-products, in which individual detector integrations are not averaged, which helps to take full advantage of the sky rotation to deconvolve the planet signal from the residual starlight. The coherent flux of the science channel were first phase-referenced to the fringe-tracker coherent flux, and then corrected for the metrology zero-point, extracted either from the HD~25535~AB observation (for the dual-field off-axis observation of 2023-11-02), or from the on-star observations (for the dual-field on-axis observations of 2023-11-02 and 2023-11-24).

\par We measured the position ($\Delta\alpha$,$\Delta\delta$) and contrast spectrum ($C_\mathrm{planet}$) of the companion relative to its bright host star from the phase-referenced complex coherent-fluxes. We used the \texttt{exogravity} pipeline\footnote{\url{https://gitlab.obspm.fr/mnowak/exogravity}}, following \citet[][]{GRAVITYCollaboration2019} and subsequent ExoGRAVITY work \citep[see in particular][]{GravityCollaboration2020, Nowak2020, Nowak2024b}.  In the absence of the stellar halo, instrumental throughput, or the Earth's atmosphere, the planet's coherent flux $\Gamma_\mathrm{planet}(b,t,\lambda)$ is described by the baseline $b$, exposure $t$, and wavelength $\lambda$:

\begin{equation}
    \Gamma_\mathrm{planet}(b,t,\lambda) = C_\mathrm{planet}(\lambda)\Gamma_\mathrm{star}(b, t, \lambda)\times e^{-i\frac{2\pi}{\lambda}(\Delta\alpha\times U(b,t) + \Delta\delta\times V(b,t))},
\end{equation}

\noindent{}where $\Gamma_\mathrm{star}$ is the stellar coherent flux, and $U(b,t)$ and $V(b,t)$ are the \textit{u}-\textit{v} plane coordinates for a given baseline and exposure time. We can model the measured coherent fluxes in the presence of a bright host, instrumental throughput, and atmospheric transmission, $\Gamma_\mathrm{on planet}(b,t,\lambda)$, as a combination of the transmission function $G$, a low order polynomial $Q$, the stellar coherent flux $\Gamma_\mathrm{star}$, and planet coherent flux $\Gamma_\mathrm{planet}$: 

\begin{equation}
    \label{eq:model}
    \Gamma_\mathrm{on planet}(b,t,\lambda) = Q(b,t,\lambda) G(b,t,\lambda) \Gamma_\mathrm{star}(b,t,\lambda) + G(b,t,\lambda) \Gamma_\mathrm{planet}(b,t,\lambda).
\end{equation}

The intermittent observations of the star using the science fiber give us a measure of $G$, $\Gamma_\mathrm{star}$, and allow us to measure $\Gamma_\mathrm{on planet}$ in terms of the contrast spectrum, the ratio of the planet and stellar spectrum, canceling out $G$ if we assume $G$ is stable across the $\Delta t$ between on planet and on star observations. In reality, the change in the atmospheric transmission across $\Delta t$ induces small telluric features in our final spectrum that result in additional correlated noise, which future work could address. 
\par These equations are non-linear in ($\Delta\alpha$,$\Delta\delta$) but not in $C_\mathrm{planet}$, so we first begin by solving for ($\Delta\alpha$,$\Delta\delta$) by assuming $C_\mathrm{planet}$ based on our reproduction of the best fitting \texttt{petitRADTRANS} model in \citet{Zhang2023}. Using the matrix form $\mathbf{\Gamma_\mathrm{on planet}}=(\Gamma_\mathrm{on planet}(b,t,\lambda))_\mathrm{b,t,\lambda}$, we define a $\chi^2$:

\begin{equation}
    \chi^2(\Delta\alpha,\Delta\delta,Q, C_\mathrm{planet}) = [\Gamma_\mathrm{on planet}-\Gamma_{\Delta\alpha,\Delta\delta,Q, C_\mathrm{planet}}]^T W^{-1} [\Gamma_\mathrm{on planet}-\Gamma_{\Delta\alpha,\Delta\delta,Q, C_\mathrm{planet}}],
\end{equation}

\noindent{}where $W$ is the covariance matrix on the projected coherent fluxes $\mathbf{\tilde{\Gamma}}$ \citep[see][Appendix A.4]{Nowak2020}. A Bayes factor is then constructed by taking $\chi^2_\mathrm{reference}=\chi^2(0,0)$, where the exponential is flat and there is no planet signal. This quantity, $z(\Delta\alpha,\Delta\delta)=\Delta\chi^2=\chi^2_\mathrm{reference}-\chi^2(\Delta\alpha,\Delta\delta)$, can be computed for any $\Delta\alpha,\Delta\delta$, but we restrict the computation to the effective field of view of the instrument. We calculate $\Delta\chi^2$ on a $200\times200$ grid about the 60 mas diameter field of view defined by the fiber's pointing and the FWHM of the fiber's throughput curve \citep[Appendix A]{Wang2021a}. Figure \ref{fig:detections} shows these detection maps in terms of the reduced $\Delta\chi^2_{\rm red}$ for the three observations, which all result in strong detections of AF~Lep~b. We take the minimum of the $\Delta\chi^2$ map as the preliminary companion position, and then calculate another $200\times200$ step grid with a range restricted to $\pm2.5\,\mathrm{mas}$ around the initial $\Delta\chi^2$ grid minimum (that is, zooming into the central ``peak" of the $\Delta\chi^2$ map). This minimum is used to initialize a gradient descent algorithm that determines the $\Delta\alpha,\Delta\delta$ best describing the companion's position given the data. The procedure is repeated for each of the individual exposures of each night. The final measurement and associated covariance matrix are taken to be the mean and covaraince of the individual measurements, following \citep{Nowak2020, Nowak2024a}. This way, the uncertainty is derived from the data itself and not, for instance, on the relatively arbitrarily defined grid upon which the $\Delta\chi^2$ maps are computed, or the uncertainty reported by the gradient descent algorithm, which may underestimate the true uncertainty. This astrometry is recorded in Table \ref{tab:astrometry}.


\begin{figure*}
    \centering
    \includegraphics[width=\textwidth]{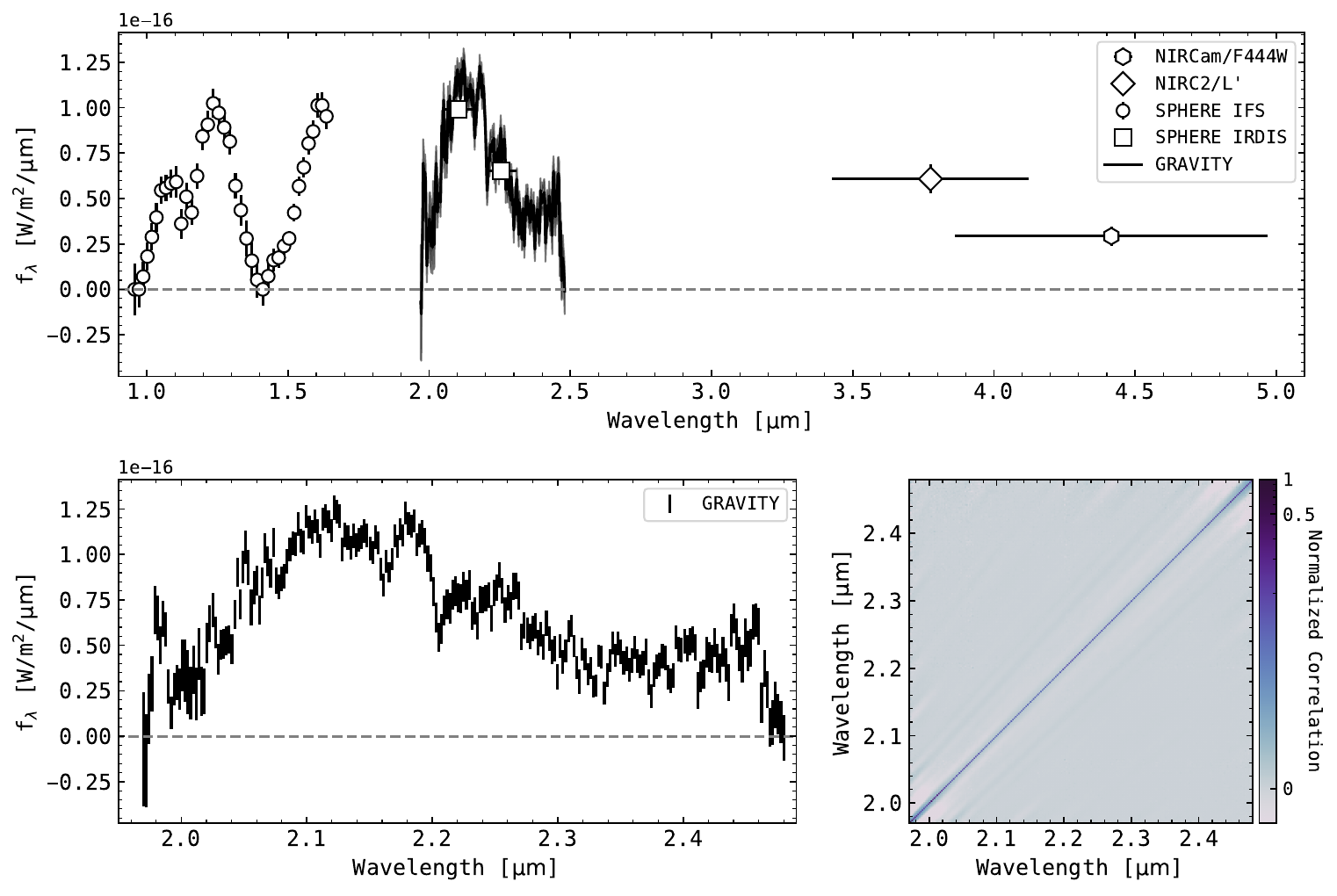}
    \caption{Spectral Energy Distribution of AF~Leporis~b in the context of new GRAVITY observations. The top panel plots our reprocessed SPHERE IFS (small circles) and IRDIS (squares) spectro-photometry (see \S\ref{subsec:previous obs}), Keck/NIRC2 Lp photometry from \citet[large circle]{Franson2023a, Franson2024}, and \textit{JWST}/NIRCam F444W photometry from \citet[hexagon]{Franson2024}, alongside the GRAVITY spectrum from this work (line). The bottom left panel plots the GRAVITY spectrum as errorbars with uncertainties corresponding to the square root of the diagonal of the covariance matrix. The spectrum is dominated by methane absorption from 2.2-2.4 microns; there is correlated noise lurking beneath the prominent molecular absorption features. The bottom right panel plots the empirical covariance matrix of the GRAVITY spectrum, in terms of the maximum value of the matrix, with a log stretch in order to illustrate the small but non-zero off-axis correlation terms. These are accounted for in all the spectral analyses in this paper. The new SPHERE and GRAVITY spectra are available as the data behind the figure.
    }
    \label{fig:spectrum}
\end{figure*}

\par One the astrometry is determined, the contrast spectrum and its covariance matrix can be extracted using a joint-fit of the coherent flux obtained on all of the baselines and individual detector integrations, holding the astrometry constant. We find that the conditions and observing strategy for the observation on 2023-11-02 result in a spectrum of significantly worse quality than the observations on 2023-11-24 and 2023-12-24; this spectrum has much larger uncertainties compared to the later two observations and does not contribute much information when the three are combined with a weighted median. We therefore use a weighted median combination of the two on-axis spectra (and their associated covariance matrices) from 2023-11-24 and 2023-12-24 for our final analysis. We transformed the contrast spectrum of the companion into a flux calibrated spectrum using a synthetic spectrum of the host star. We scaled a \texttt{BT-NextGen} \citep{Allard2011} spectrum with $T_{\rm eff}=6000\,\mathrm{K}$, $\log(g)=4.3$, and $\mathrm{[Fe/H]}=-0.27$ \citep{Zhang2023} to archival photometry from \textit{Gaia} \citep{GaiaCollaboration2022}, Tycho2 \citep{Hog2000}, and 2MASS \citep{Cutri2003} using \texttt{species} \citep{Stolker2020}. We sampled this scaled model spectrum on the GRAVITY wavelength grid using \texttt{spectres} \citep{Carnall2017}. 
Our final flux calibrated $K$-band spectrum of AF~Lep~b is shown in Figure \ref{fig:spectrum}, alongside previous observations. We find that our spectrum agrees well with the observed SPHERE K1 and K2 photometry (see next section).

\begin{deluxetable*}{ccccccccccc}
\tablewidth{\textwidth}
\tablecaption{Observing log. Science Camera (SC) and Fringe Tracker (FT) targets and exposures are recorded. NEXP, NDIT, and DIT denote the number of exposures, the number of detector integrations per exposure, and the detector integration time, respectively. $\tau_0$ denotes the atmospheric coherence time. The fiber pointing is the placement of the science fiber relative to the fringe tracking fiber (which is placed on the central star), $\gamma$ is the coupling efficiency at the position of the companion, depending on the distance between pointing and planet location (see Table \ref{tab:astrometry}). \label{tab:obslog}}
\tablehead{Date & \multicolumn{2}{c}{Target} & \multicolumn{2}{c}{NEXP/NDIT/DIT} & Airmass & $\tau_0$ & Seeing & Fiber pointing & $\gamma$ \\
\colhead{[UT]} & SC & FT & SC & FT & & & & $\Delta$RA/$\Delta$DEC &}
\startdata
\hline\hline
2023-11-02 & HD~25535~A/B & HD~25535~B/A & 4/64/1~s & 4/64/1~s & 1.03-1.06 & 2.4-3.0~ms & $1.09-1.40^{\prime\prime}$ & 718.32/803.6 & 0.999  \\
2023-11-02 & AF~Lep~b & AF~Lep~A & 14/4/100~s & 2/64/1~s & 1.03-1.09 & 1.6-2.4~ms & $1.12-1.92^{\prime\prime}$ & 316.5/64.3  & 0.999 \\
2023-11-24 & AF~Lep~b & AF~Lep~A & 18/12/30~s & 5/16/3~s & 1.03-1.15 & 4.8-13.8~ms & $0.41-0.76^{\prime\prime}$ & 319.71/57.54 & 0.988 \\
2023-12-24 & AF~Lep~b & AF~Lep~A & 23/12/30~s & 4/48/1~s & 1.03-1.31 & 6.4-11.5~ms & $0.41-0.68^{\prime\prime}$ & 315.38/55.84  & 0.999
\enddata
\end{deluxetable*}

\begin{deluxetable*}{cCCCCCC}
\tablewidth{\textwidth}
\tablecaption{Relative astrometry of AF~Lep~b from VLTI/GRAVITY. \label{tab:astrometry}}
\tablehead{
\colhead{Date} & \colhead{Epoch} & \colhead{$\Delta$RA} & \colhead{$\sigma_{\Delta\alpha}$} & \colhead{$\Delta$Dec} & \colhead{$\sigma_{\Delta\delta}$} & \colhead{$\rho$} \\
\colhead{[UT]} & \colhead{[MJD]} & \colhead{[mas]} & \colhead{[mas]} & \colhead{[mas]} & \colhead{[mas]} &  \colhead{}
}
\startdata
2023-11-02 & 60251.32 &  316.89 & 0.07 & 62.91 & 0.09 & -0.73 \\ 
2023-11-24 & 60273.22 &  316.26 & 0.04 & 60.17 & 0.04 & -0.82 \\ 
2023-12-24 & 60302.21 &  315.44 & 0.05 & 55.93 & 0.05 & -0.55
\enddata
\tablecomments{The co-variance matrix for each measurement can be reconstructed using $\sigma_{\Delta\alpha}^2$ and $\sigma_{\Delta\delta}^2$ on the diagonal, and $\rho\times\sigma_{\Delta\alpha}\times\sigma_{\Delta\delta}$ on the off-diagonal.}
\end{deluxetable*}

\subsection{Previous observations} \label{subsec:previous obs}

\par For our spectral analysis, we re-reduced the two pupil tracking, angular differential imaging (ADI) mode observations of AF~Lep taken with the VLT/SPHERE coronagraphic direct imaging instrument on 2022-10-16 and 2022-12-20 that were presented in \cite{Mesa2023} to achieve a higher signal-to-noise ratio (SNR) spectrum by leveraging the \texttt{PACO} algorithm. The re-reduction was computed on the COBREX Data Center, a modified and improved server based on the High Contrast Data Center, \citep[HC-DC, formerly known as the SPHERE Data Center;][]{Delorme_sphereDC_2017}. Since the SPHERE observations on 2022-10-20 were taken in the star hopping reference differential imaging (RDI) mode \citep{Wahhaj2021, DeRosa2023} and have a significantly lower integration time and the higher airmass than the ADI observations, we do not re-reduce them or use them in our spectral analysis, as that would involve fitting flux (and potentially wavelength-calibration) offset parameters during the atmospheric modeling step \citep{Zhang2023, Palma-Bifani2024}.
The pre-reduction steps consisting of dark, flat, distortion, and bad pixel corrections are based on the SPHERE data reduction and handling (DRH) pipeline provided by ESO \citep{Pavlov_DRH_2008}. A few additional customs step were added for the IFS prereduction to the DRH, mainly to correct the cross talk during the spectral extraction and improve the bad pixel correction \citep{Mesa_ifs_2015}.

We post-processed the data using the \texttt{robust PACO ASDI} algorithm \citep{Flasseur_paco_2018, flasseur_robustness_2020, Flasseur_asdi_2020}.
\texttt{PACO} estimates the nuisance component using a multi-Gaussian model at a local scale on small patches, allowing for a better estimation of the temporal and spectral correlation of the nuisance.
Details on the improvements of the pre-reduction pipeline, the optimization regarding the ASDI mode of \texttt{PACO} and the obtained performances can be found in \cite{Chomez2023a}. \texttt{PACO} provides a contrast gain between one and two magnitudes at all separations as well as reliable and statistically grounded SNR detection and contrast maps in an unsupervised and data-driven fashion. 
The two extracted spectra are averaged to produce the spectrum shown in Figure \ref{fig:spectrum}.

\par In our spectral analysis we also include a weighted average of the Keck/NIRC2 \textit{L'} photometry from \citet{Franson2023c, Franson2024} and the \textit{JWST}/NIRCam F444W photometry from \citet{Franson2024}.

\section{Analysis} \label{sec:analysis}

\subsection{Orbital analysis} \label{subsec:orbit}

\par In order to determine the architecture of the AF~Lep system we fit the observations with two-body Keplerian orbits using \texttt{orvara} \citep{Brandt2021}. This code was chosen because it has been predominately used in previous studies of the system \citep{DeRosa2023, Franson2023c, Mesa2023, Zhang2023, Bonse2024}. The orbit fits included our new GRAVITY observations, the literature relative astrometry available prior to our observations\footnote{The two epochs of SPHERE astrometry from \citet{Mesa2023}, 
the single epoch of SPHERE astrometry from \citet{DeRosa2023}, the Keck/NIRC2 \textit{L'} astrometry from \citet{Franson2023c}, and the recently recovered archival VLT/NACO \textit{L'} astrometry from \citet{Bonse2024}.}, the absolute astrometry of the system from the HGCA, and the HIRES radial velocity observations of the star \citep{Butler2017}. 
\par
\begin{figure*}

\gridline{
\fig{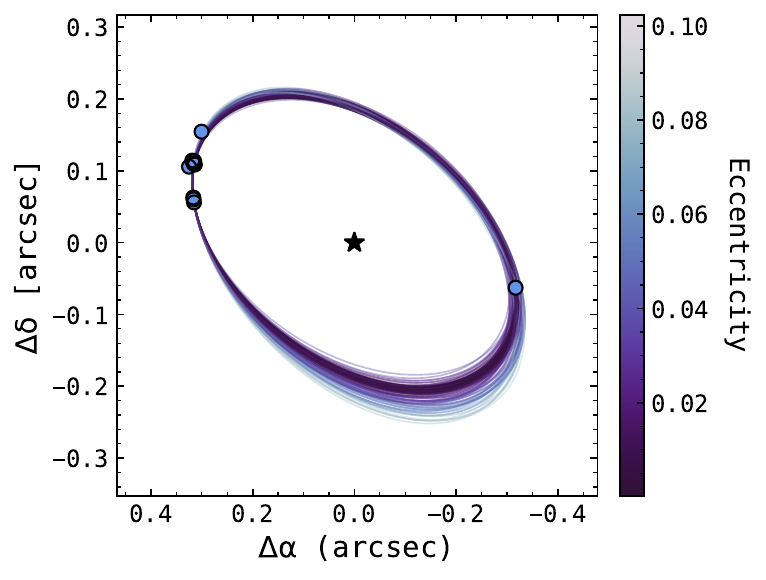}{0.6\textwidth}{(a)}
}
\gridline{
\centering
\fig{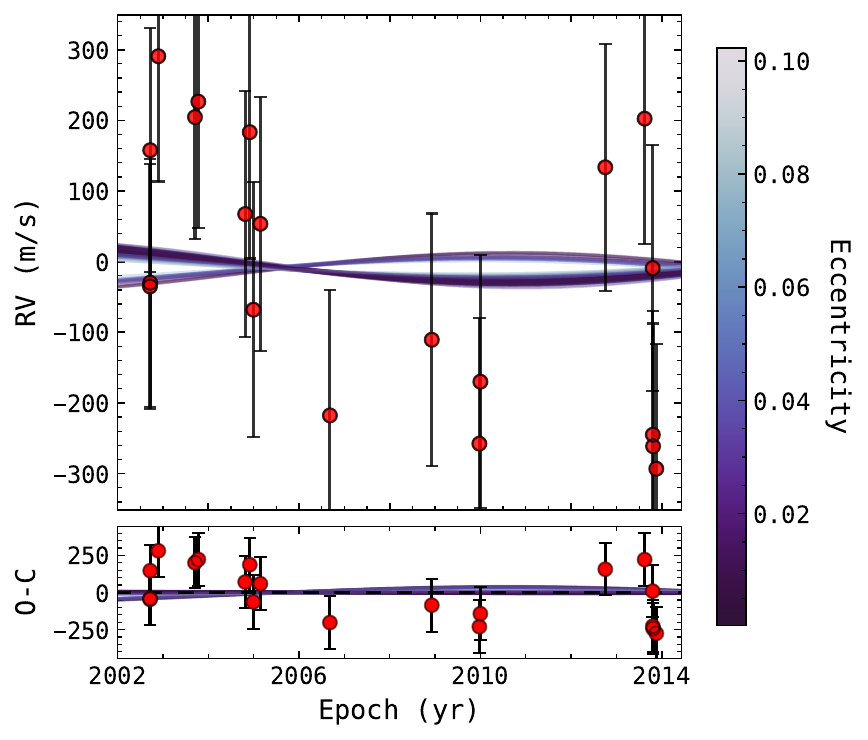}{0.3\textwidth}{(b)}
\fig{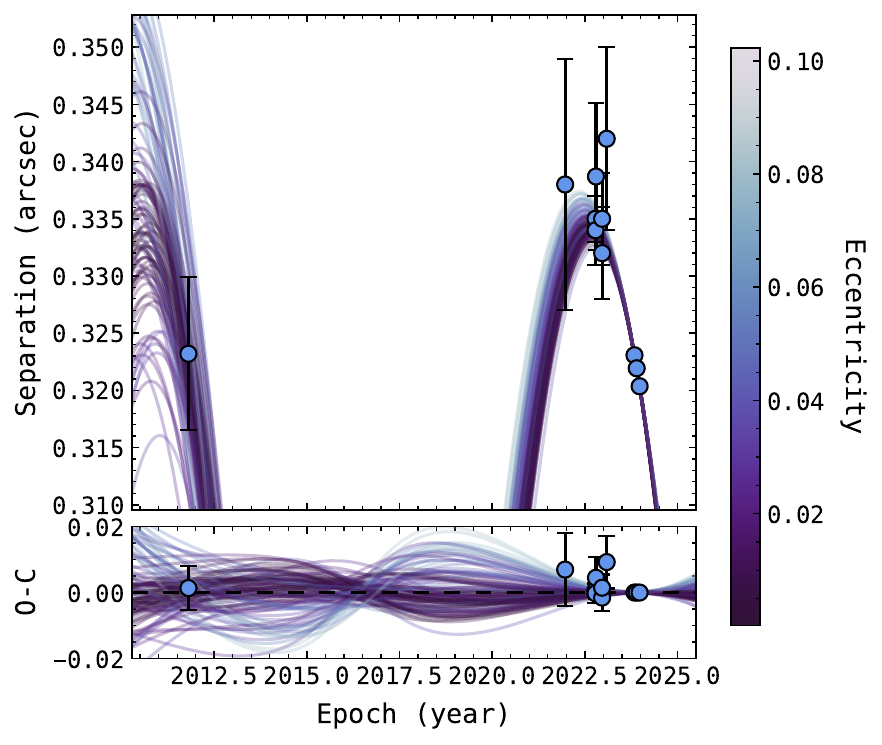}{0.3\textwidth}{(c)}
\fig{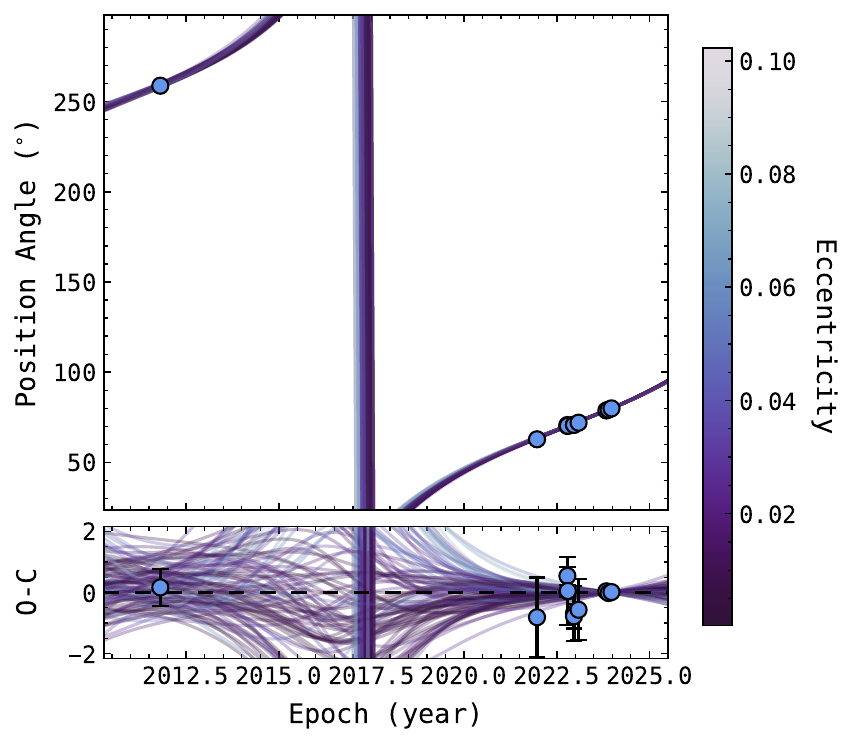}{0.3\textwidth}{(d)}
}
\gridline{
\centering
\fig{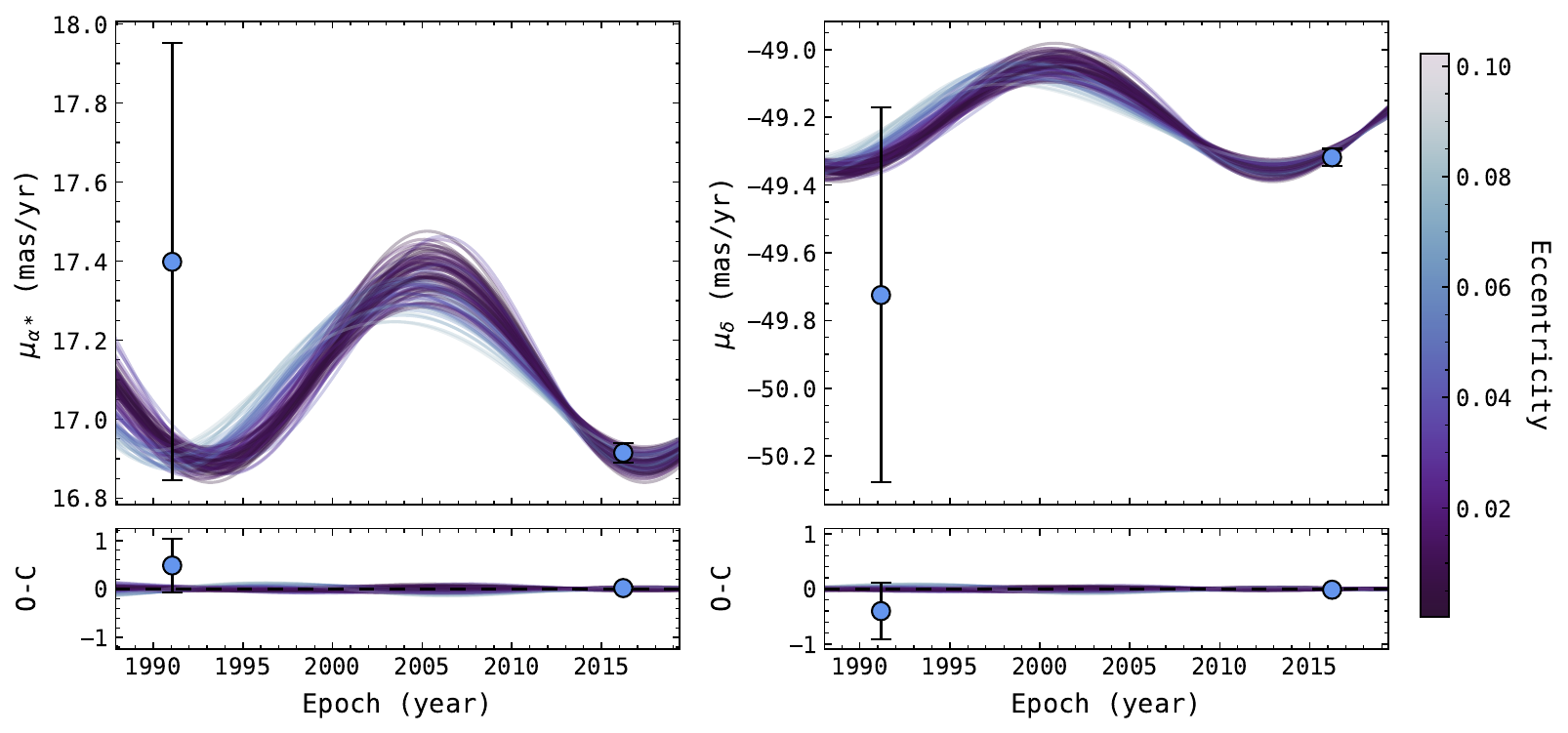}{0.6\textwidth}{(e, f)}
}
\caption{Two-body orbits fit to the observations of the AF~Lep system. (a) Sky-projected relative astrometry of AF~Lep~b (blue scatter points) and fit orbits (lines, color coded by eccentricity). The new high precision VLTI/GRAVITY relative astrometry constrains the eccentricity and inclination of the orbit, and when coupled with the long time baseline of the VLT/NACO recovery \citep{Bonse2024}, results in a well known, circular orbit. (b) Keck/HIRES radial velocity measurements of AF~Lep~A and the best fitting orbits. The available radial velocities do not constrain the orbit fit, and there remain two degenerate solutions for the argument of periastron ($\omega$) and longitude of ascending node ($\Omega$). (c) Projected separation versus time for each orbit, and relative astrometry. There is larger scatter between single mirror instruments, but these are consistent within $1\,\sigma$ with the best fitting circular orbits that are driven by the GRAVITY astrometry. (d) Likewise, for the position angle versus time for each orbit and the relative astrometry. (e, f) The proper motions in R.A. and Dec. for AF~Lep~A (blue scatter points) and the reflex motion due to the orbit of AF~Lep~b (colored lines).}
\label{fig:orbit}
\end{figure*}

\begin{figure}
    \centering
    \includegraphics[width=0.75\linewidth]{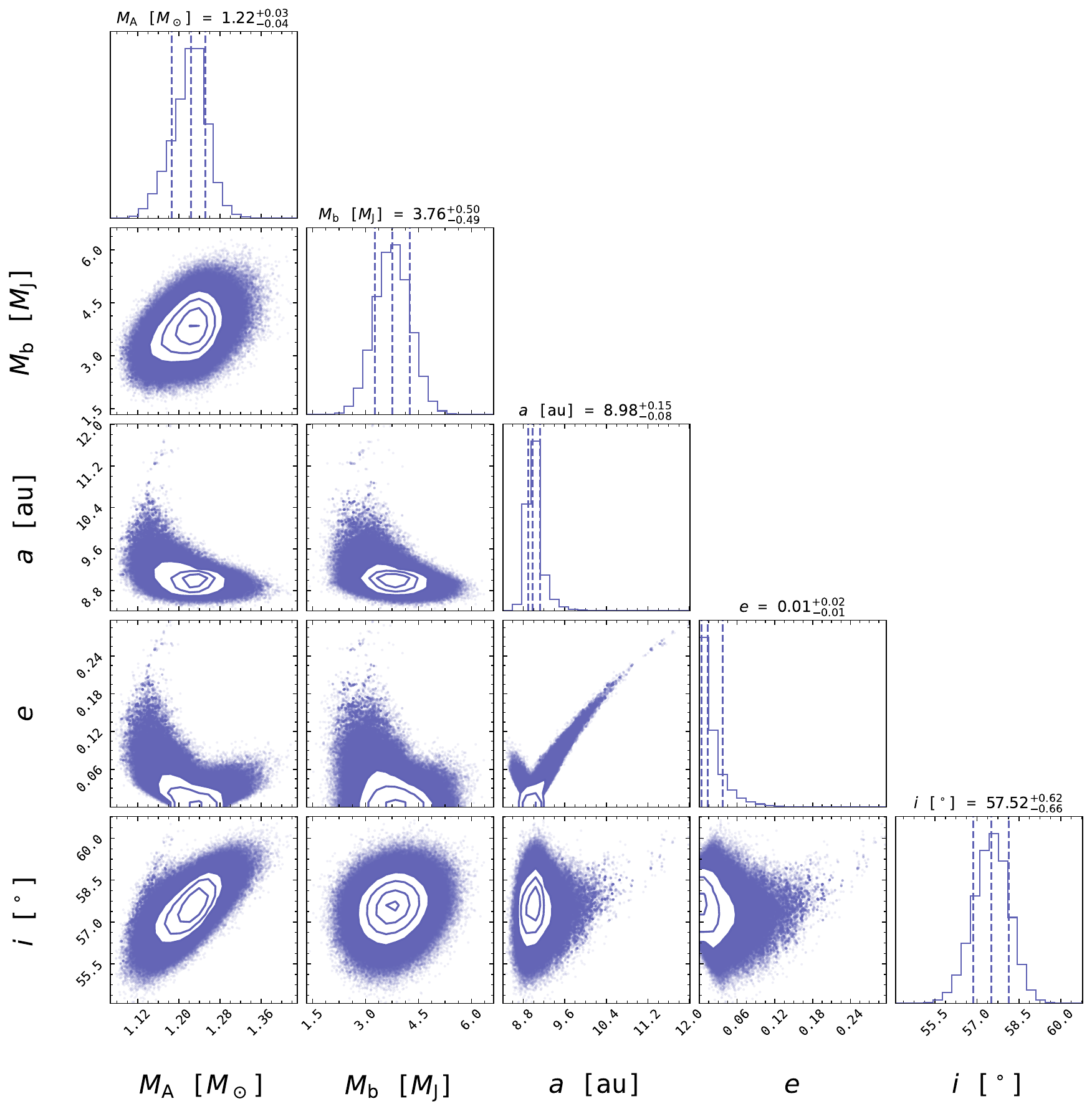}
    \caption{Posterior distribution of select parameters from the \texttt{orvara} orbit fit visualized in Figure \ref{fig:orbit} and summarized in Table \ref{tab:orbit}. Complex correlations are apparent between $a$ and $e$ that result in a low probability tail of more eccentric orbits. The majority of accepted orbits are circular. The component masses are well constrained, roughly normal distributions.}
    \label{fig:orbit_post}
\end{figure}

 Our \texttt{orvara} runs use the \texttt{ptemcee} sampler \citep{Foreman-Mackey2013, Vousden2016} with 20 temperatures, 1000 walkers, 500,000 steps per walker, and were thinned to only save every 100th step; we visually inspected the chains for convergence and discarded the first 3000 saved steps (so the first 60\% of each chain) as ``burn-in". We fit for the mass of AF~Lep~A, the mass of AF~Lep~b, the orbit semi-major axis ($a$), parameterized forms of the eccentricity and argument of periastron ($\sqrt{e}\cos(\omega_\star)$, $\sqrt{e}\sin(\omega_\star)$), the inclination angle ($i$), the longitude of the ascending node ($\Omega$), the mean longitude at the reference epoch 2010-01-01 ($\lambda_\mathrm{ref}^*$), the system parallax ($\varpi$), and RV zeropoint and jitter terms. We adopt uninformative priors on all parameters \citep[see][]{Brandt2021}, except for the system parallax (we adopt the \textit{Gaia} DR3 parallax as a Gaussian prior) and the primary mass \citep[we adopt $\mathcal{N}(1.20,0.06)\,M_\odot$,][]{Franson2023c}. The $\sqrt{e}\cos(\omega_\star)$, $\sqrt{e}\sin(\omega_\star)$ eccentricity and argument of periastron parameterization is used to address a known bias in sampling circular orbits with eccentricity as a free parameter \citep{Lucy1971, Zakamska2011}.  Table \ref{tab:orbit} records the results of this fit, while Figure \ref{fig:orbit} illustrates them. The posterior distribution of key parameters is shown in Figure \ref{fig:orbit_post}. 
\par This updated fit represents a significant improvement in the constraints on the planet's orbit. We note here that the majority of previous work has been unable to accurately constrain the circular nature of the planet's orbit given the precision of the data and limited coverage of the planet's orbital arc since its discovery.  This resulted in orbit fits with posterior distributions of eccentricity dominated by the uniform prior distribution---this effect has been detailed most recently in \citep{Blunt2023}---, with $e$ ranging from $\sim0.5$ \citep{Mesa2023} to $\sim0.2$ \citep{Franson2023c}. The recovery of the planet in archival VLT/NACO data \citep{Bonse2024} recently facilitated a likelihood dominated orbit fit, which indicated an upper limit on $e<0.06, 0.15, 0.26$, at $1, 2, 3\,\sigma$, respectively. Compared to this previous orbit fit, our fit sets the limit at $e<0.02, 0.07, 0.13$ at $1, 2, 3\,\sigma$, respectively. We discuss the implications of these constraints in \S\ref{subsec:discuss_orbit}.

\begin{deluxetable*}{lll}
\tablewidth{0pt}
\tablecaption{Orbital analysis of AF~Lep~A+b with \texttt{orvara} \label{tab:orbit}}
\tablehead{
        {\bf Parameter} & 
        {\bf Prior} &
        {\bf Median and $1\,\sigma$ CI}
        }
\startdata
\multicolumn{3}{c}{Fitted parameters} \\ \hline
Primary Mass $M_{\rm pri}$  ($M_{\odot}$)      &  $\mathcal{N}(1.20, 0.06)$ & ${1.22}_{-0.04}^{+0.03}$ \\
Secondary Mass $M_{\rm sec}$  ($M_{J}$)    &  $1/M$ (log-flat) & $3.75\pm0.5$ \\
Semimajor axis $a$  (au)              &  $1/a$ (log-flat) & ${8.98}_{-0.08}^{+0.15}$ \\
$\sqrt{e} \sin \omega$  &  uniform & $-0.03\pm0.12$ \\
$\sqrt{e} \cos \omega$  &  uniform & ${0.00}_{-0.08}^{+0.08}$ \\
Inclination $i$  ($^\circ$)                 & $\sin(i)$, 0$^{\circ}<i<180 ^{\circ}$ & ${57.5}_{-0.7}^{+0.6}$ \\
Mean longitude at 2010.0 $\lambda_{\rm ref}^*$ ($^\circ$)  &  uniform & ${171.8}_{-1.6}^{+2.1}$ \\
Ascending node $\Omega$  ($^\circ$)         &  uniform & $68.8_{-0.7}^{{+0.4}^\dag}$ \\
Parallax $\varpi$ ($\mathrm{mas}$)                &  $\mathcal{N}$(\textit{Gaia} DR3) & ${37.254}\pm0.019$ \\
System proper motion in R.A. $\mu_\alpha$ (mas ${\rm yr}^{-1}$) & uniform & $17.13\pm0.01$ \\
System proper motion in Dec. $\mu_\delta$ (mas ${\rm yr}^{-1}$) & uniform & $-49.20\pm0.01$ \\
RV Zero-point RV$_{0}$  $({\rm m\,s}^{-1})$ & uniform & $-10\pm2$ \\
RV Jitter  $\sigma_{\rm jit}$ $({\rm m\,s}^{-1})$ &  $1/\sigma$ (log-flat) & ${175}_{-29}^{+38}$ \\
\hline
\multicolumn{3}{c}{Derived parameters}\\
\hline
Period (years)                                    & \nodata &   ${24.3}_{-0.4}^{+0.9}$ \\
Argument of periastron $\omega\, (^{\circ})$      & \nodata &   ${289}_{-21}^{{+13}^\dag}$  \\
Eccentricity $e$                        & \nodata &   ${0.013}_{-0.010}^{+0.024}$ $^\ddag$   \\
Semimajor axis $a$ (mas)                          & \nodata &  ${334.4}_{-2.9}^{+5.6}$  \\
Time of periastron $T_{0} = \mathrm{t_{ref}} - P\frac{\lambda - \omega}{360^{\circ}}$ (JD) & \nodata & ${24580}_{-1300}^{+4200}$ \\
Mass ratio                                        & \nodata &   ${0.00294}_{-0.00036}^{+0.00036}$
\enddata
\tablenotetext{*}{The reference epoch is $\mathrm{t_{ref} = 2455197.5\, JD\, (\lambda_{ref}; 2010\, Jan\, 1\, 00:00\, UT)}$. }
\tablenotetext{\dag}{Degenerate solution, with values $\pm180^\circ$, so there exists a second population of orbits with $\Omega{=}248_{-0.7}^{{+0.4}^\circ}$, $\omega{=}109_{-21}^{{+13}^\circ}$, $\lambda_{\rm ref}^*{=}351.8_{-1.6}^{{+2.1}^\circ}$ in the posterior distribution.}
\tablenotetext{\ddag}{The marginalized posterior distribution on eccentricity encompasses 0.0. This is effectively an upper-limit, $e<0.02, 0.07, 0.13$ at $1, 2, 3\,\sigma$.}
\end{deluxetable*}

\subsection{Atmospheric analysis} \label{sec:spec_analysis}

\subsubsection{Evolutionary model predictions} \label{subsec:evol_models}

\par AF~Lep~b has been and will continue to be studied intently in the context of evolutionary modeling as its parameters are refined \citep{DeRosa2023, Franson2023c, Mesa2023, Zhang2023, Gratton2024, Palma-Bifani2024}. Within the year since its discovery, the specific estimates of its bulk properties as predicted by evolutionary models has varied widely, because of the systematic uncertainty in the estimates of its dynamical mass (see \S\ref{subsec:discuss_orbit}). There are also systematic uncertainties in the age estimate for the $\beta$-Pictoris moving group \citep[see discussion in][]{Lee2024}, but given the generally adopted isochronal age of $24\pm3\,\mathrm{Myr}$ \citep{Bell2015}, previous work has converged on a rough range of bulk parameters for the planet, namely effective temperatures $700-850\,\mathrm{K}$, surface gravities of $\log(g)\sim3.7$ and radii of about $R\sim1.3\,R_{\rm J}$. In general, we expect that regardless of initial entropy the planet's radius will not be smaller than $1.1\,R_{\rm J}$ at an age less than a few hundred million years \citep{Marley2012}, and certainly no less than about $0.9\,R_{\rm J}$ at any age \citep[e.g., Figure 3 in][]{Burrows2001}. As the goal of this paper is to assess primarily the orbit and atmosphere of the planet in the context of its formation, we limit our evolutionary model analysis to that which will help contextualize our atmospheric analysis \citep[see Table 3 and Figure 3 in][that consider a wide variety of evolutionary models---we adopt their evolutionary model based priors in the sections below]{Zhang2023}. We used \texttt{species} to linearly interpolate the hybrid-cloudy evolutionary model grid from \citet{Saumon2008} to establish a general range of expected planetary effective temperatures, surface gravities, radii, and luminosities given the dynamical mass and isochronal moving group age. We find $M_{\rm b}=3.75\pm0.5\,M_{\rm J}$ and $24\pm3\,\mathrm{Myr}$ implies $T_{\rm eff}=770\pm75$, $\log(g)=3.73\pm0.06$, $R=1.30\pm0.01\,R_{\rm J}$, and $\log(L/L_\odot)=-5.26\pm0.18$. These are recorded in Table \ref{tab:atmo_analysis}.

\subsubsection{Self-consistent atmospheric models} \label{subsec:forward_model_setup}

\par We perform initial reconnaissance of the atmospheric properties of AF~Lep~b using pre-computed grids of self-consistent radiative-convective-equilibrium (RCE) models. We compared the GRAVITY spectrum to the modern, cloudless \texttt{Sonora Bobcat} \citep{Marley2021} and \texttt{Sonora Elf Owl} \citep{Mukherjee2024} grids; the former being a solar abundance, rainout chemical equilibrium model and the latter varying abundances in terms of metallicity and C/O ratio, in addition to including the effects of disequilibrium chemistry by transport induced mixing by parameterizing the Eddy diffusion coefficient $\log(K_{\rm zz})$. Neither of these cloudless models reproduce the Y- or J-band flux of the object, they predict bluer spectra than are observed. We compared the full spectral energy distribution (SED) of the object to the cloudy, solar abundance \texttt{BT-Settl} model grid \citep{Allard2011}, which resulted in a poor fit to the SPHERE and GRAIVTY spectra. We primarily focus our discussion on the \texttt{Exo-REM} forward model grid \citep{Baudino2015, Charnay2018}. This model parameterizes non-equilibirum chemistry between C, N, and O bearing molecules with profiles of the eddy diffusion coefficient that vary with pressure determined for each grid point, and parameterizes the cloud particle distribution using a simplified microphysical treatment that produces cloud particle radii intermediate to the fixed radius or fixed sedimentation parameter case. The publicly available grid\footnote{\url{https://lesia.obspm.fr/exorem/YGP_grids/}} varies temperature ($T_\mathrm{eff}$), surface gravity ($\log(g)$), carbon-to-oxygen ratio (C/O), and metallicity compared to Solar ([M/H]). The grid varies temperature between $400$ and $2000\,\mathrm{K}$ in steps of $50\,\mathrm{K}$, surface gravity between 3.5 and 5.0 in steps of 0.5, [M/H] between -0.5 and 2 (0.3 to 100 times solar) with steps of 0.5, and C/O between 0.1 and 0.8 in steps of 0.05. A previous version of this grid was used by \citet{Palma-Bifani2024} who found that the existing SPHERE datasets and \textit{L'}-band photometry were best fit by enriched metallicity ($>0.4$) models; their grid did not include the highest metallicity grid points ($\mathrm{[M/H]}\in[1,2]$), however. The updated version of the grid we use here encompasses the high metallicity solutions found in the retrieval analysis in \citet{Zhang2023, Franson2024}.

\begin{figure}
    \centering
    \includegraphics[width=\textwidth]{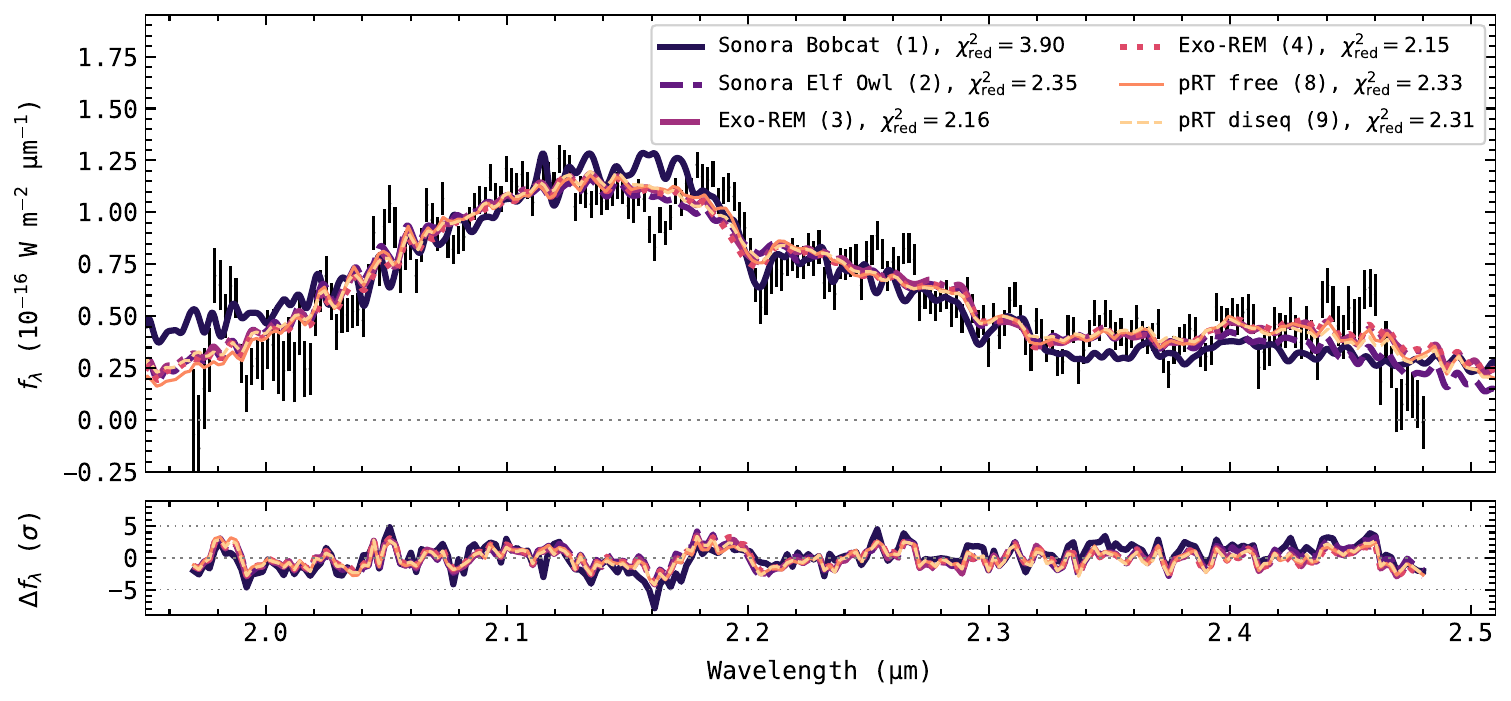}
    \caption{Model comparisons to the GRAVITY spectrum (not including additional datasets). Shown are models, numbered based on their appearance in Table \ref{tab:atmo_analysis}, and the reduced $\chi^2$ of the fit. 
    Most models provide a reasonable fit to the spectrum, but only the cloudy \texttt{Exo-REM} grid and the cloudy, free chemistry \texttt{petitRADTRANS} retrieval result in physically plausible radii ($R>1.1\,R_{\rm J}$).}
    \label{fig:gravonly_spectrum_fits}
\end{figure}

\begin{figure}
    \centering
    \includegraphics[width=\textwidth]{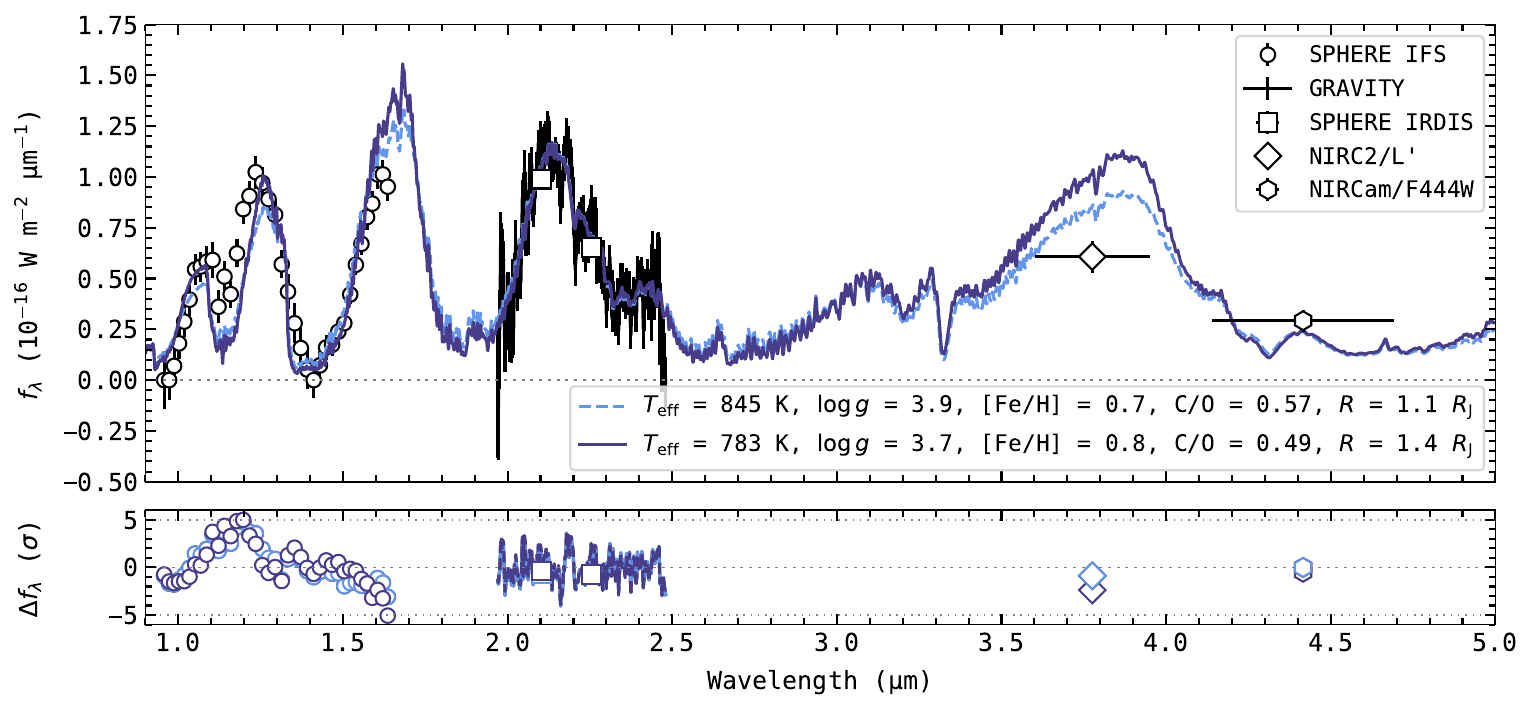}
    \caption{Interpolated \texttt{Exo-REM} atmospheric models assuming the dynamical mass as a prior (dashed light blue) and fixing the dynamical mass and evolutionary model prediction for radius (solid dark blue). To better fit the SPHERE H1 band spectrophotometry and the Keck/NIRC2 \textit{L'} photometry, the sampler prefers unphysically small radii (high $\log(g)$), higher effective temperatures, higher C/O, and lower metallicity. The GRAVITY data rules out higher temperature models, very low and very high C/O ratios, and solar metallicities, and is matched well by the model in either case.}
    \label{fig:fm_spectrum_fits}
\end{figure}

\par We use \texttt{species} to linearly interpolate the spectra between grid points and compare our data to the model. In our fitting routine, we sample the grid parameters, parallax (which determines the absolute flux of the model along with planetary radius), and, motivated by the presence of residual correlated noise in our SPHERE spectrum \citep[e.g.][]{Greco2016}, a Gaussian process parameterized by a squared-exponential kernel as an estimate for said correlated noise \citep[see Equation 4 in][]{jaWang2020}. The empirical GRAVITY correlation matrix (computed in \S\ref{subsec:gravity}) is included in our likelihood function. We used \texttt{pyMultiNest}\footnote{\href{https://johannesbuchner.github.io/PyMultiNest/}{johannesbuchner.github.io/PyMultiNest/}} \citep{Feroz2008, Feroz2009, Buchner2014} to sample the interpolated grid with 500 live points. We place a Gaussian prior on the parallax based on the \textit{Gaia} DR3 measurement \citep{GaiaCollaboration2022}. Following \citet{Palma-Bifani2024}, when fitting the \texttt{Exo-REM} grid we attempted to place a prior on $\log(g)\sim3.7$ informed by the dynamical mass and evolutionary model predictions for the radius, but we found that any prior that reasonably captured the systematic uncertainty in the evolutionary radius (e.g. $\pm0.1\,\mathrm{dex}$) was always driven away from the prior. We therefore compare posteriors fixing the $\log(g)=3.7$ and mass to $M_{\rm b}=3.75\,M_\mathrm{Jup}$ versus setting a Gaussian prior equivalent to our dynamical mass estimate on the mass ($M_{\rm b}=3.75\pm0.5\,M_\mathrm{Jup}$) and letting radius and $\log(g)$ vary freely. 

\par Table \ref{tab:atmo_analysis} records the results from this grid interpolation fit comparison. Figure \ref{fig:gravonly_spectrum_fits} compares our new GRAVITY spectrum to the \texttt{Sonora} models, \texttt{Exo-REM} model, and \texttt{petitRADTRANS} agnostic cloud retrievals (see next section). Figure \ref{fig:fm_spectrum_fits} compares the results of the \texttt{Exo-REM} models fit to the full dataset. The results of these model comparisons are discussed in Section \ref{subsec:discuss_atmo}.

\begin{rotatepage}
\movetabledown=2.6in
\begin{table*}
\footnotesize
\centering
\begin{rotatetable}
\caption{Subset of parameters for AF~Lep~b derived from evolutionary and atmospheric analysis.}
\label{tab:atmo_analysis}
\begin{tabular}{lccccccccc}
\hline\hline
\multicolumn{1}{c}{Model} & \multicolumn{8}{c}{Notable Model Parameters}  & $\chi_\mathrm{red.}^2$/d.o.f. \\
 & Mass & $T_{\rm eff}$ & log(g) & Radius & [Fe/H] & C/O & $\log(K_{\rm zz})$/$P_{\rm quench}$ $^*$ & $\log(L/L_\odot)$ & \\
\multicolumn{1}{c}{} &  [$M_{\rm J}$] & [$\mathrm{K}$] &  & [$R_{\rm J}$] &  &  & [$\log(\rm cm^2/s)$]/[$\mathrm{bar}$] &  & \\
\hline
\multicolumn{1}{c}{Evolutionary} & & & & & & & & & \\
SM08 (hybrid-cloudy), $24\pm3\,\mathrm{Myr}$ & $3.75\pm0.5$ & $770\pm75$ & $3.73\pm0.06$ & $1.30\pm0.01$ & solar & solar & \nodata & $-5.26\pm0.18$ & \nodata \\
\hline
\multicolumn{1}{c}{Self-consistent RCE grid interpolation$^\dag$} & & & & & & \\
\multicolumn{1}{c}{--- Gravity spectrum only ---} & & & & & & & & \\

\texttt{Sonora Bobcat}, $\mathcal{U}(0.5, 2.0)_\mathrm{R_p}$ & $0.18\pm0.02$ & $1260\pm4$ & $3.25\pm0.01$ & $0.50^{+0.01}_{...}$ $^\ddag$ & $0.22\pm0.02$ & \nodata & \nodata & $-5.22\pm0.01$ & 3.90/228 \\

\texttt{Sonora Elf Owl}, $\mathcal{U}(0.5, 2.0)_\mathrm{R_p}$ & $3.02\pm0.51$ & $1080\pm50$ & $4.05\pm0.15$ & $0.82\pm0.09$ & $0.05\pm0.03$ & $0.3^{+0.04}_{...}$ $^\ddag$ & $7.73\pm0.69$ & $-5.06\pm0.02$ & 2.35/226 \\

\texttt{Exo-REM}, $\mathcal{U}(0.5, 1.25)_\mathrm{R_p}$$^\S$ & $3.53\pm0.55$ & $836^{+7.5}_{-10}$ & $3.80\pm0.10$ & $1.17^{+0.05}_{-0.03}$ & $0.63\pm0.05$ & $0.54\pm0.03$ & \nodata & $-5.20\pm0.01$ & 2.16/227 \\

\texttt{Exo-REM}, $\mathcal{U}(1.25, 2.0)_\mathrm{R_p}$$^\S$ & $3.80\pm0.45$ & $762^{+16}_{-10}$ & $3.65^{+0.06}_{-0.08}$ & $1.46^{+0.07}_{-0.05}$ & $0.66\pm0.06$ & $0.48^{0.02}_{-0.04}$ & \nodata & $-5.16\pm0.02$ & 2.15/227 \\

\multicolumn{1}{c}{--- All data ---} & & & & & & & & & \\

\texttt{BT-Settl}, $\mathcal{U}(0.5, 2.0)_\mathrm{R_p}$ & $4.21\pm0.04$ & $766\pm2$ & $3.50^{+0.01}_{...}$ $^\ddag$ & $1.81\pm0.01$ & solar & solar & \nodata & $-4.97\pm0.01$ & 6.02/270 \\

\texttt{Exo-REM}, $\mathcal{U}(0.5, 2.0)_\mathrm{R_p}$ & $4.05\pm0.46$ & $845\pm3$ & $3.89\pm0.05$ & $1.13\pm0.01$ & $0.69\pm0.02$ & $0.57\pm0.01$ & \nodata & $-5.21\pm0.01$ & 2.45/268 \\

\texttt{Exo-REM}, $M_{\rm b}=3.75\,M_\mathrm{Jup}$, $\log(g)=3.7$ & $=3.75$ & $783\pm5$ & $=3.7$ & $1.35\pm0.01$ & $0.77\pm0.03$ & $0.49\pm0.01$ & \nodata & $-5.18\pm0.01$ & 2.63/266 \\

\hline

\multicolumn{1}{c}{\texttt{petitRADTRANS} atmospheric retrieval$^\dag$} & & & & & & & \\
\multicolumn{1}{c}{--- Gravity spectrum only ---} & & & & & & & \\
Agno., Grad. P-T, free chem., $\mathcal{U}(0.5, 2.0)_\mathrm{R_p}$ & $3.6\pm0.4$ & $851\pm40$ & $3.78\pm0.14$ & $1.22^{+0.19}_{-0.14}$ & \nodata & $0.43\pm0.05$ & \nodata & $-5.12^{+0.05}_{0.08}$ & 2.33/204  \\

Agno., Grad. P-T, (dis)eq. chem., $\mathcal{U}(0.5, 2.0)_\mathrm{R_p}$ & $3.65\pm0.4$ & $1113\pm50$ & $4.27\pm0.08$ & $0.70\pm0.05$ & $0.40\pm0.11$ & $0.50\pm0.05$ & $2.67\pm0.06$ & $-5.15\pm0.03$ & 2.31/212  \\ 

\multicolumn{1}{c}{--- All data ---} & & & & & & & & \\

Cloudless, spline P-T, (dis)eq. chem., $\mathcal{U}(0.5, 2.0)_\mathrm{R_p}$ & $3.9\pm0.4$ & $947\pm20$ & $4.22\pm0.05$ & $0.76\pm0.03$ & $0.83\pm0.08$ & $0.61\pm0.04$ & $2.59^{+0.10}_{-0.14}$ & $-5.33\pm0.01$ & 2.49/247  \\ 

Agno., Grad. P-T, free chem., $\mathcal{U}(0.5, 2.0)_\mathrm{R_p}$ & $3.7\pm0.4$ & $980\pm25$ & $4.22\pm0.06$ & $0.74\pm0.03$ & \nodata & $0.56\pm0.05$ & \nodata & $-5.32\pm0.01$ & 2.57/245 \\ 

EddySed, 3-part P-T, (dis)eq. chem., $\mathcal{U}(0.5, 2.0)_\mathrm{R_p}$ & $3.9\pm0.4$ & $970\pm14$ & $4.23\pm0.06$ & $0.75\pm0.03$ & $0.79\pm0.08$ & $0.53\pm0.04$ & $2.28\pm0.22$ & $-5.33\pm0.02$ & 2.57/253  \\ 

EddySed, Grad. P-T, free chem., $\mathcal{U}(0.5, 2.0)_\mathrm{R_p}$ & $3.8\pm0.3$ & $972\pm24$ & $4.22\pm0.06$ & $0.75\pm0.04$ & \nodata & $0.52\pm0.04$ & \nodata & $-5.31\pm0.02$ & 2.47/244  \\ 

EddySed, Grad. P-T, (dis)eq. chem., $\mathcal{U}(0.5, 2.0)_\mathrm{R_p}$ & $3.8\pm0.4$ & $947\pm19$ & $4.17\pm0.06$ & $0.80\pm0.04$ & $0.74\pm0.09$ & $0.50\pm0.04$ & $2.84\pm0.05$ & $-5.32\pm0.01$ & 2.42/252  \\ 

EddySed, Grad. P-T, (dis)eq. chem., $\mathcal{U}(1.2, 1.5)_\mathrm{R_p}$ & $4.1\pm0.3$ & $789\pm9$ & $3.82\pm0.04$ & $1.23^{+0.02}_{...}$ $^\ddag$& $0.78\pm0.08$ & $0.61\pm0.04$ & $2.64\pm0.05$ & $-5.25\pm0.02$ & 2.51/252  \\ 


EddySed, Grad. P-T, (dis)eq. chem., $\mathcal{N}(1.33,0.03)_\mathrm{R_p}$ & $4.0\pm0.4$ & $790^{+5}_{-10}$ & $3.83\pm0.04$ & $1.23^{+0.02}_{...}$ $^\ddag$ & $0.75\pm0.07$ & $0.60\pm0.04$ & $2.64\pm0.05$ & $-5.25\pm0.01$ & 3.06/246 \\ 

\hline

Adopted values & $3.75\pm0.5$ & $800\pm50$ & $3.7\pm0.2$ & $1.3\pm0.15$ & $0.75\pm0.25$ & $0.55\pm0.10$ & $2.6^{+0.2}_{-0.3}$ & $-5.2_{-0.2}^{+0.1}$ & \nodata \\ 

\hline
\end{tabular}
\tablecomments{For each type of model considered (evolutionary, self-consistent atmosphere, atmospheric retrieval) we record the mean and $1\,\sigma$ confidence interval for the parameters of interest. $\mathcal{U}(x,y)_\mathrm{P}$ denotes a uniform from $x$ to $y$, and $\mathcal{N}(x,y)_\mathrm{P}$ denotes a normally distributed prior with mean $x$ and standard deviation $y$ on the parameter P. There are 233+39+4-$\Sigma P_i$=276-$\Sigma P_i$ degrees of freedom for each comparison considering all the data visualized in Figure \ref{fig:spectrum}, and 233-$\Sigma P_i$ for each comparison considering only the GRAVITY spectrum. $^*$\,We present $\log(K_{\rm zz})$ for RCE grids that include it as a free parameter, and $P_{\rm quench}$ for \texttt{petitRADTRANS} retrievals that implement (dis)equilibrium chemistry. $^\dag$\,All model fits use our dynamical mass estimate as a prior influencing $\log(g)$ given $R$. $^\ddag$\,Denotes a parameter that has piled up at a prior boundary. $^\S$\,When fitting only the GRAVITY spectrum with linear interpolations of the \texttt{Exo-REM} grid, we observed a multimodal posterior; we report the median and CI for the low and high temperature modes in two lines in this table. 
}
\end{rotatetable}
\end{table*}
\end{rotatepage}

\subsubsection{An initial analysis using atmospheric retrievals} \label{subsec:retrieval_setup}

\par We investigated the atmospheric composition of AF~Lep~b by iteratively sampling thousands of highly parameterized atmospheric models and comparing the resulting spectra to our data; this commonly referred to as an atmospheric ``retrieval" \citep[for a review, see][]{Madhusudhan2018}. In this framework, we parameterize the pressure-temperature structure of the atmosphere, the chemical abundances, clouds, and bulk atmospheric properties in order to facilitate the rapid calculation of spectra and sample the posterior distribution of parameters. The drawback to this technique is that these parameterized model atmospheres are not necessarily physically consistent, but the technique allows for the flexible exploration of many parameters of interest and can indicate missing physics via comparison with other retreival permutations or with RCE models. 
\par We used the \texttt{petitRADTRANS} radiative transfer package to model the emission spectrum of AF~Lep~b and to conduct our atmospheric retrievals \citep{Molliere2019, Molliere2020, Alei2022}. We refer the reader to \citet{Molliere2020, Zhang2023, Nasedkin2024} for full descriptions of the retrieval implementation, particularly the chemistry. We also refer the reader to \citet{Nasedkin2023a} for a description of the code and sampler, and to \citet{Molliere2020, Nowak2020, Kammerer2021, Balmer2023} for previous usage of the retrieval in the context of GRAVITY observations. \texttt{species} was used as a wrapper for setting up the model framework and for initializing nested sampling via \texttt{pyMultiNest}, now with 2000 live points in constant sampling efficiency mode with an efficiency of 0.05 in order to explore the posterior distributions. 
\par To briefly summarize, these retrievals use the correlated-k treatment for opacities, including line species for H$_2$O \citep{ExoMol_H2O}, CO \citep{rothman_hitemp_2010}, CH$_4$ \citep{ExoMol_CH4}, CO$_2$ \citep{ExoMol_CO2}, NH$_{3}$ \citep{ExoMol_NH3}, HCN \citep{ExoMol_HCN}, H$_{2}$S \citep{ExoMol_H2S}, PH$_{3}$ \citep{exomol_ph3}, FeH \citep{wende_feh_2010}, Na \citep{allard_new_2019}, K \citep{allard_k-h_2016}, SiO \citep{exomol_sio}, TiO \citep{exomol_tio}, and VO \citep{exomol_vo}, Rayleigh scattering opacities for H$_2$ and He \citep{Dalgarno1962, Chan1965}, and collision induced absorption (CIA) of H$_2$ and He \citep{Borysow1988, Borysow1989, Borysow2001, Richard2012}. Many of these opacities are available from the ExoMol database \citep{Chubb2021}, in the ``petitRADTRANS" format.\footnote{\href{https://www.exomol.com/data/data-types/opacity/}{www.exomol.com}}
\par \citet{Nasedkin2023b} showed that unaccounted for correlations in input data for spectral retrievals can strongly bias the resulting atmospheric inferences. As in \S\ref{subsec:forward_model_setup}, we estimate the correlated noise in the SPHERE IFS spectrum using a Gaussian process by sampling the parameters of a squared exponential kernel that estimates the correlation matrix of the spectrum. Again, we provide the retrieval likelihood function the empirical GRAVITY correlation matrix. 
\par We considered two chemistry parameterizations, both described in \citet{Nasedkin2024}. The first is based on an initial assumption of equilibrium chemistry modified by transport-induced quenching. Here, metallicity [M/H] and C/O are free parameters and abundances of individual molecules at a given pressure and temperature (set by the retrieved P-T profile) are determined by interpolating along a temperature/pressure/[M/H]/C/O table that was computed using \texttt{easyCHEM} \citep{molliere2017}. To approximate the effect of transport-induced chemical disequilibrium (``vertical mixing"), a ``quench pressure" parameter is sampled. Above this presssure, the abundances of H$_2$O, CO, and CH$_4$ are set constant to the value at $P_{\rm quench}$. We refer to this as the (dis)equilibrium chemistry paramterization. The second chemistry parameterization is refered to as ``free" chemistry, where the vertically constant abundances for each molecule are sampled as free parameters, with the constraint that the sum of these mass fractions is less than one.
\par Three pressure-temperature paramterizations were considered, and are described in \citet{Nasedkin2024}. One retrieval relied on a cubic ``spline" interpolation profile with an oscillation penalty term \citep{Line2015}. Another used the ``3-part" parameterization introduced in \citet{Molliere2020}, which samples three free temperature nodes above the photosphere, follows the Eddington profile in the photosphere, and is forced onto a moist adiabat up to the radiative convective boundary. The majority of the retrieval experiments used the ``gradient" parameterization introduced by \citet{Zhang2023}, where the gradient of the P-T profile is sampled, instead of the temperature at a given pressure. This parameterization allows for prior constraints on the gradient based on ``self-consistent" RCE models. The gradient $d\ln T/d\ln P$ is sampled at six pressure nodes spaced logarithmically between $1000-10^{-3}\,\mathrm{bar}$, and normally distributed priors are adopted for each node based on the results of a sample of self-consistent radiative-convective equilibrium models. We adopt the same priors on $d\ln T/d\ln P$ listed in Equation 3 of \citet{Zhang2023} and Table 6 in \citet{Nasedkin2024}, derived from the RCE models of \citet{Phillips2020, Marley2021, Karalidi2021, Mukherjee2022, Lacy2023}.

\par The majority of our retrievals used the \citet{Ackerman2001} parameterization for clouds \citep[see \S2.4 in][for the implementation of this cloud model in \texttt{petitRADTRANS}]{Molliere2020}, which is referred to as the ``EddySed" model here. This model defines a cloud sedimentation efficiency parameter $f_{\rm sed}$ that sets the cloud mass fraction $X^c$, $$X^c(P)=X_0^c\left(\frac{P}{P_{\rm base}}\right)^{f_{\rm sed}}$$ where $P_{\rm base}$ is found by intersecting our P-T profile with the saturation vapor pressure profile of a given cloud species. $X_0^c$, the value of $X^c$ at $P_{\rm base}$, is allowed to vary in terms of $X_{\rm eq}^c$, the cloud mass fraction assuming equilibrium condensation at $P_{\rm base}$. The cloud particle vertical eddy diffusion coefficient $K_{zz}$ is also allowed to vary, which effectively sets an average particle size given $f_{\rm sed}$. The width of a log-normal particle size distribution $\sigma_g$ about this average particle size is another free parameter. We included opacity contributions from iron \citep[Fe,][]{Henning1996}, enstatite \citep[MgSiO$_3$,][]{Scott1996, Jaeger1998}, and potassium chloride \citep[KCl,][]{Barton2014} grains. The crystalline (DHS irregular) opacities were adopted because the observations do not reach long enough wavelengths to distinguish between amorphous and crystalline grains for any of the species considered. These are calculated using OpacityTool \citep{Min2005}, which makes use of software from \citet{Toon1981}. The wavelength dependent opacities are added to the wavelength bins, as they vary slowly compared to the line opacities.
\par The fully parameteric cloud model, introduced as ``cloud model 2" in \citet{Molliere2020}, that we dub the ``agnostic" cloud model was used when conducting retreivals on only the GRAVITY spectrum, since the clouds considered here are ``gray," that is, wavelength independent across the \textit{K}-band. This model has 
\begin{equation}
    \kappa_{\rm tot} = \kappa(P)\left(\frac{\lambda}{\lambda_R}\right)^{\xi}
\end{equation}
\noindent{} where $\kappa_{\rm tot}$ is the total cloud opacity (both scattering and absorbing components), $\kappa(P)$ is the value at $1\mu\rm m$ at pressure $P$, $\lambda_R$ is a reference wavelength, and $\xi$ is the spectral slope of the cloud. Then, we define $\kappa(P)$
\begin{equation}
\kappa_{\rm cloud}(\lambda, P)= \kappa_0 \left(\frac{P}{P_{\rm base}}\right)^{f_{\rm sed}},
\end{equation}
\noindent{} where $P<P_{\rm base}$, where $\kappa_0$, $f_{\rm sed}$ are free parameters describing the cloud base pressure in units of bar and the cloud opacity in units of $\mathrm{cm^2/g}$ at the reference wavelength, and the cloud scale height die off. We also parameterize a single scattering albedo $\omega$ such that
\begin{equation}
    \kappa_{\rm abs} = (1-\omega)\kappa_{\rm tot}.
\end{equation}
\noindent{} This model was also briefly compared against the full dataset, coupled with the gradient P-T profile, but was not preferred when compared to the EddySed model.
\par
\begin{figure}
    \centering
    \includegraphics[width=\textwidth]{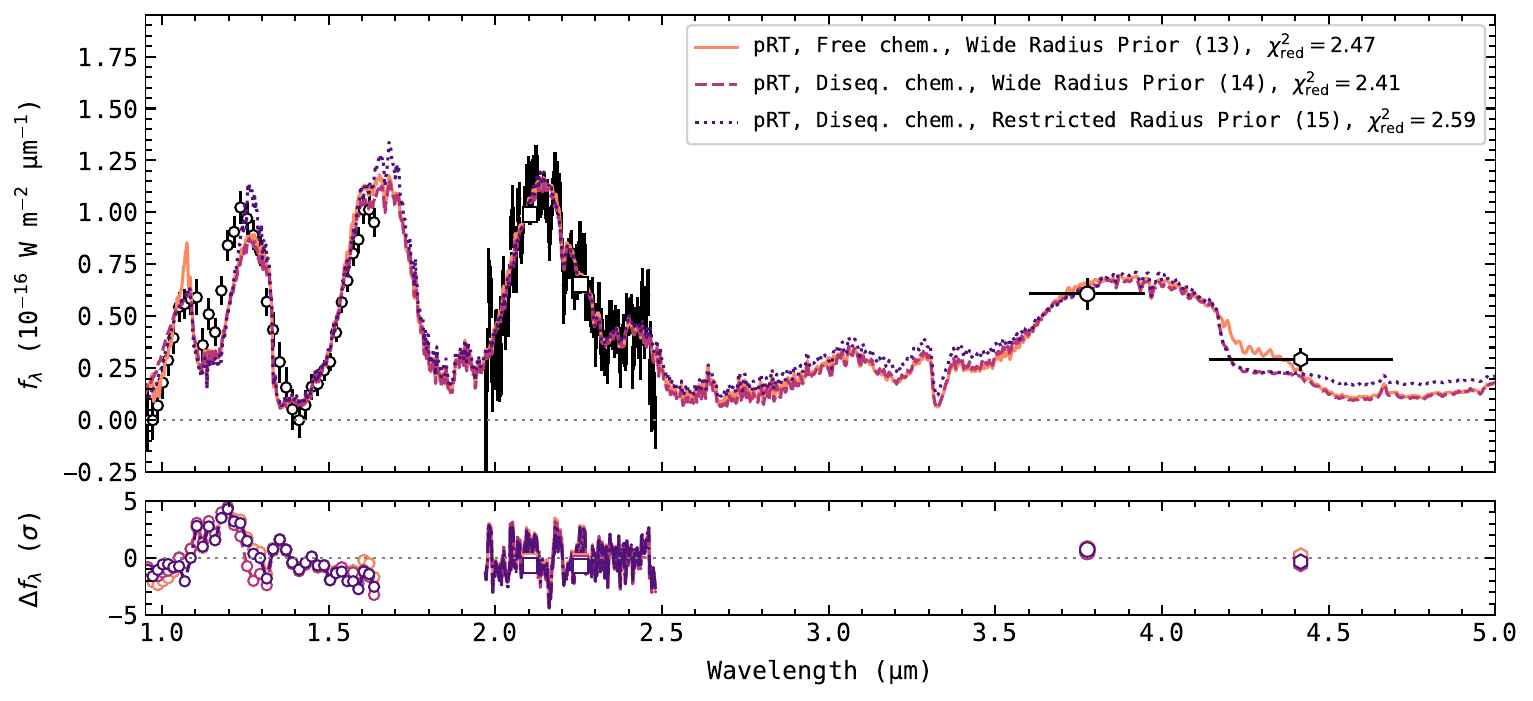}
    \caption{\texttt{petitRADTRANS} retrieval examples, as described in \S\ref{subsec:retrieval_setup}. These three models use the EddySed cloud model with Fe, MgSiO$_3$, KCl clouds and the RCE-prior-constrained gradient P-T profile. The solid orange line represents a retrieval run with free chemistry and a wide (0.5-2.0$\,R_{\rm J}$) uniform prior on radius; the dashed magenta line, a run with (dis)equilibrium chemistry and the same wide uniform radius prior; the dark purple dotted line, a run with (dis)equilibrium chemistry and a  restricted (1.2-1.5$\,R_{\rm J}$) uniform radius prior following \citet{Zhang2023}. Each retrieval is labeled with its reduced $\chi^2$, their associated parameters appear in Table \ref{tab:atmo_analysis}, lines 13-15.}
    \label{fig:petitradtrans_spectra}
\end{figure}

\begin{figure}
    \centering
    \includegraphics[width=\textwidth]{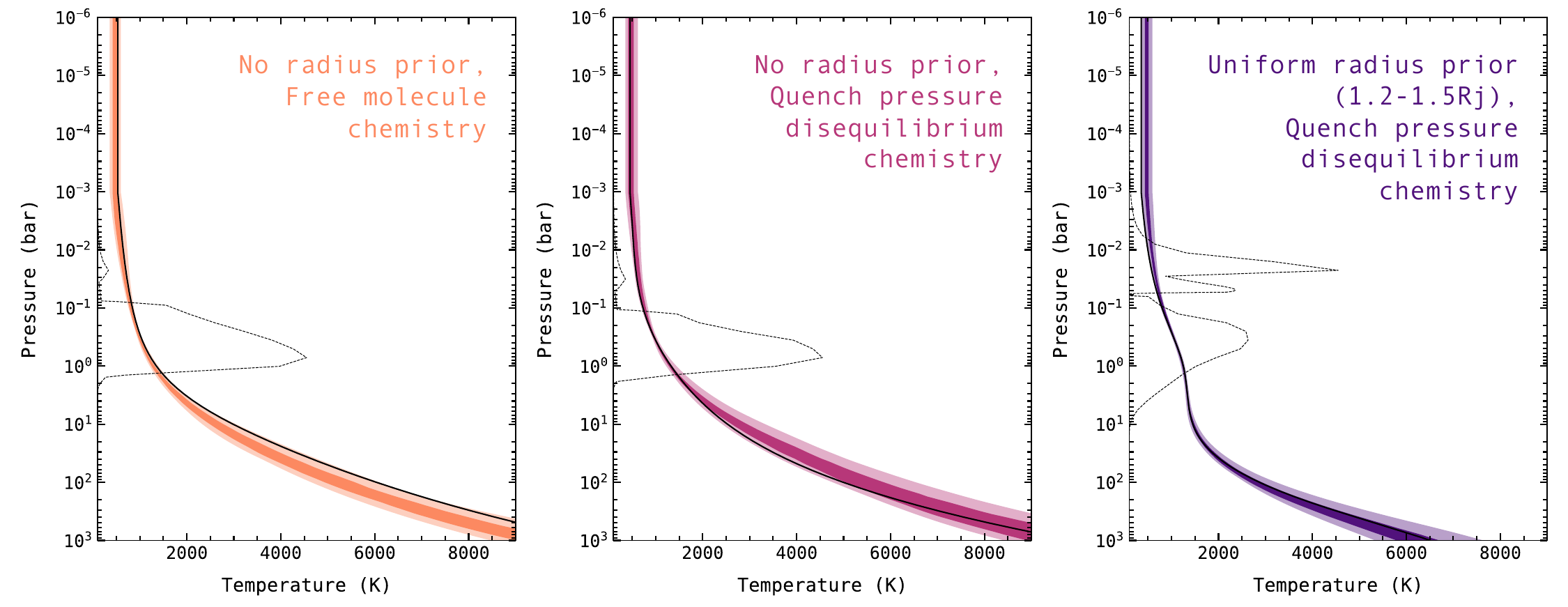}
    \caption{Pressure-temperature profiles corresponding to the retrievals shown in Figure \ref{fig:petitradtrans_spectra}. When the sampler is forced to accept solutions with $R>1.2\,R_{\rm J}$, the P-T profile is forced away from radiative convective equilibrium and into an isothermal shape. This could be indicative of missing cloud/haze opacity due to the parameterization of cloud particle sizes, as discussed in \S\ref{subsec:discuss_atmo}.}
    \label{fig:petitradtrans_pt}
\end{figure}

 We constructed a series of retrieval tests driven by outstanding questions raised in \citet{Zhang2023} and \citet{Palma-Bifani2024}; these are recorded in Table \ref{tab:atmo_analysis} and discussed in \S\ref{sec:discuss}. These experiments were designed to answer 1) how does the new GRAVITY spectrum change, revise, or confound the results of previous studies, 2) what do the composition inferences suggest about the planet's properties and formation history, 3) what data and model deficiencies are present, and 4) what information is expressed in the GRAVITY \textit{K}-band spectrum of an ``AF~Lep~b-like" planet, that is, what can one learn from retrievals conducted on the GRAVITY spectrum alone? The results are discussed regarding points 1-3 in \S\ref{subsec:discuss_atmo}-\ref{subsec:discuss_formation} and 4) in \S\ref{subsec:discuss_grav+gaia}. The ``GRAVITY-only" retrievals are included in Figure \ref{fig:gravonly_spectrum_fits}, and a subset of the full SED retrievals are visualized in Figure \ref{fig:petitradtrans_spectra}, and their corresponding P-T profiles are shown in \ref{fig:petitradtrans_pt}. No retrieval resulted in a derived radius that was consistent with evolutionary model expectations within uncertainties, aside from the free chemistry ``GRAVITY-only" retrieval that has a large uncertainty ($R=1.22^{+0.19}_{-0.14}\,R_{\rm J}$) and fails to predict the observed \textit{Y}- and \textit{J}-band. 

\section{Discussion} \label{sec:discuss}

\subsection{The orbit of AF~Lep~b: formation, system architecture, and debris disk stirring} \label{subsec:discuss_orbit}

\par Recent population level studies of directly imaged planets indicate that ``planetary mass" ($\lesssim13\,M_\mathrm{Jup}$) companions exhibit on average lower eccentricity than brown dwarfs \citep{Bowler2020, Nagpal2023, DoO2023}, although a larger sample and longer orbital coverage is likely needed to confirm this \citep[see discussion in ][]{DoO2023}. Radial velocity surveys, with a larger sample size skewed towards significantly shorter periods than directly imaged companions, show that the eccentricity distribution of single gas giants appears to pile up at circular orbits with a long tail out to $e\sim0.8$ \citep{Rosenthal2024}. Both of these results suggest that formation within a gas disk, where drag forces can damp initial eccentricity, result in populations of planets with low eccentricity. Our updated eccentricity measurement, an upper limit of $e<0.02, 0.07, 0.13$ at $1, 2, 3\,\sigma$, is therefore consistent with expectations based on formation within a protoplanetary disk. Our updated inclination angle measurement, $i_{\rm orbit}={57.5}_{-0.7}^{{+0.6}^\circ}$, is consistent with the stellar inclination angle $i_{\star}=54_{-9}^{{+11}^\circ}$ \citep{Franson2023c}, and therefore with a system in spin-orbit alignment\footnote{Because the stellar polar position angle is unknown, this apparent agreement only represents a lower-limit on the true obliquity angle \citep[\S2]{Bowler2023}.}. Spin-orbit alignment, like circular orbits, is preferentially set by formation within the protoplanetary disk, and there is growing evidence for spin-orbit alignment of directly imaged planets when compared against brown dwarf companions \citep{Bowler2023, Sepulveda2024}.
\par The precisely constrained semi-major axis of the planet, $a=8.98^{+0.15}_{-0.08}\,\mathrm{au}$, is also coincident with the peak of the recently observed radial velocity giant planet occurrence rate maximum \citep{Fulton2021, Lagrange2023}, which has been explained by invoking preferential formation of giant planets near their parent protoplanetary disk's ice lines. 
\par The dynamical mass measurement, $M_{\rm b}=3.75\pm0.5\,M_{\rm J}$, is consistent with that found by previous studies, and the differences in central value are attributable to the correlation between the dynamical mass on the eccentricity (see, Figure \ref{fig:orbit}, panel \textit{e} and \textit{f}). The value is just shy of the $\sim4\,M_{\rm J}$ population split indicated by radial velocity surveys \citep{Santos2017, Schlaufman2018}, which has been suggested to be due to different formation for objects above (gravitation instability) and below (core accretion) the boundary. This precise and now systematically robust dynamical mass estimate also enables the placement of AF~Lep~b in the planetary mass-metallicity relationship, as discussed below, and provides a useful prior on the surface gravity of the planet for the atmospheric analysis.
\par This updated orbit also constrains the degree to which AF~Lep~b can excite (or ``stir") planetessimals in the system, enhancing the dust production in the unresolved debris belt. The planet's discovery already ruled out the planet's ability to directly ``sculpt" the debris disk \citep{Franson2023c}. The stirring argument is interesting insofar as recent theoretical work has cast doubt on the ability of debris disks to ``self-stir," indicating that excitation via secular processes from planets might be required to reproduce the population of observed debris disks \citep{Krivov2018, Pearce2022}. Following \citet{Franson2023c}, by inverting Equation 15 from \citet{Mustill2009}, we can estimate that the minimum mass necessary for a planet on AF~Lep~b's orbit to stir the exterior debris disk. \citet{Mustill2009} give

\begin{equation}
    \frac{m_{\rm pl}}{M_\odot} = 1.53\times10^3 \left(\frac{1\,\mathrm{yr}}{t_{\rm cross}} \right) \frac{(1-e_{\rm pl})^{(3/2)}}{e_{\rm pl}} \left(\frac{a_{\rm disk}}{10\,\mathrm{au}}\right)^{(9/2)} \left(\frac{a_{\rm pl}}{1\,\mathrm{au}} \right) \left(\frac{m_\star}{M_\odot}\right)^{(1/2)}. 
\end{equation}

\begin{figure}
    \centering
    \includegraphics[width=0.65\linewidth]{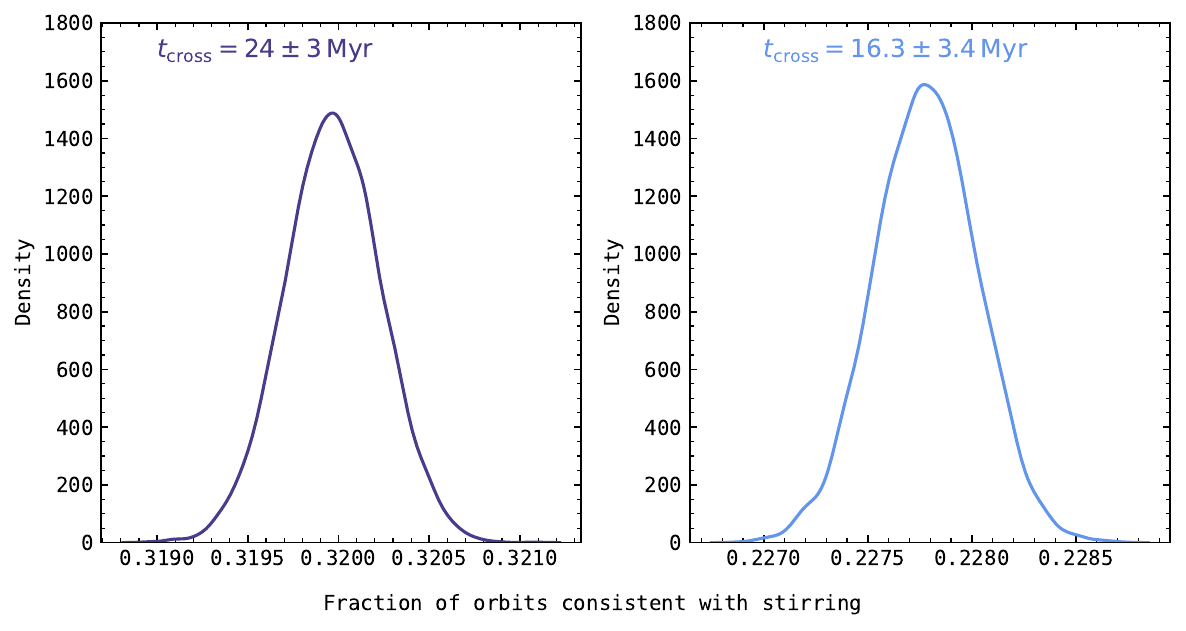}
    \caption{Fraction of AF~Lep~b orbits consistent with stirring AF~Lep's debris disk at $46\pm9\,\mathrm{au}$ \citep[following][]{Mustill2009, Pearce2022} for (left) the isochronal age \citep{Bell2015} and (right) the kinematic traceback age \citep{Lee2024}.}
    \label{fig:stirring}
\end{figure}

If we then take $t_{\rm cross}$ as the age of the system, $a_{\rm disk}=46\pm9\,\mathrm{au}$ \citep{Pearce2022}, and then adopt our posterior distribution on the orbital parameters, we can solve for $m_{\rm pl}$ and compare this to the dynamical mass constraint from the same orbit. The maximum a-posteriori orbit, with $e=0.02$, $a=8.98\,\mathrm{au}$, $m_\star=1.22\,M_\odot$, the minimum mass for stirring given $t_{\rm cross}=24\,\mathrm{Myr}$ is $m_{\rm pl}=4.9\,M_{\rm J}$ (or, for $t_{\rm cross}=16.3\,\mathrm{Myr}$, $m_{\rm pl}=7.2\,M_{\rm J}$) versus the dynamical mass of $M_{\rm b}=3.75\,M_{\rm J}$. So, our most probable orbit is inconsistent with generating the dust in the system, which would require additional sculpting or stirring planets further out to explain. Given the deep imaging limits from \textit{JWST}/NIRCam \citep{Franson2024}, such a planet would be of sub-Saturn mass. Considering the entire posterior distribution of orbits, we randomly sample from a normal distribution of $a_{\rm disk}$ and $t_{\rm cross}$, and calculate the fraction of the posterior orbits that are consistent or inconsistent with stirring the disk. We performed this calculation 5,000 times for both the isochronal system age $24\pm3\,\mathrm{Myr}$ \citep{Bell2015} and the kinematic traceback age $16.3\pm3.4\,\mathrm{Myr}$ \citep{Lee2024}. The result (Figure \ref{fig:stirring}) was that $32\%$ of the orbits are consistent with stirring a planetessimal population at $46\pm9\,\mathrm{au}$ if $t_{\rm cross}=24\pm3\,\mathrm{Myr}$, and that $23\%$ are consistent with stirring that same population if $t_{\rm cross}=16.3\pm3.4\,\mathrm{Myr}$. Largely, these are orbits in the tails of the posterior, with small but non-zero eccentricities. 
\par AF~Lep~b, with a nearly circular eccentricity but a relatively large mass, could be the sole planet responsible for ``stirring" the system's debris. This seems a simple, plausible explanation given the data. However, given that 2/3 of the posterior are inconsistent with stirring, an equally plausible explanation is that smaller, undetected planets (like sub-Neptunes or super-Earths) at wider separations are shepherding the debris more directly, analogous to Jupiter and Neptune's influence on the Kuiper belt. In the near future, continued monitoring of the system at high precision can strengthen this constraint regardless of the systematic uncertainty on the system age. Constraints on the distribution of the debris could be gained by observations with ALMA or \textit{JWST}/MIRI. It will be interesting to understand going forward whether other, less massive bodies (that would be consistent with the non-detection from \textit{JWST}/NIRCam) are needed to shape the debris in the system.

\subsection{``The radius problem"} \label{subsec:small-radius-context}

At a population level, a ``small radius problem" arises for field (mature, ages $>1$ Gyrs) brown dwarfs at the L-T (and T-Y) transitions, where atmospheric models struggle to reproduce radii estimated using evolutionary tracks, when both classes of tools are fit to match near infrared observations \citep[see Figure 24 in][]{Sanghi2023}. The small radius problem remains a persistent issue for retrievals, even when the dynamical mass is known, or mid-infrared observations are available, or high resolution observations are available \citep{Burningham2017, Zalesky2019, Gonzales2020, Kitzmann2020, Burningham2021, Zhang2021a, Zhang2021b, Gonzales2022, Lueber2022, Xuan2022, Balmer2023, Franson2023a, Hood2023, Vos2023, Xuan2024b}. There is also an equivalent but opposite issue, a ``large radius problem”, for hotter M-type objects: evolutionary models predict smaller radii and colder temperatures than are measured spectroscopically. In \citet{Balmer2024}, we found agreement at the $1-2\,\sigma$ level when forward modeling of a young, low mass ``benchmark" M-type companion, highlighting the variance of this problem: individual targets and observations may be more or less affected by this problem depending on a number of factors, like spectral type, data coverage, resolution, or model choice. Both small and large radius problems affect retrieval studies as well as forward modeling studies \citep[e.g.][in the case of retrievals on M-dwarf companions]{Xuan2024a}. This issue is understood to be an atmospheric model deficiency, rather than an evolutionary model deficiency, since it would be physically impossible for an object of order $10^{0}-10^{2}\,M_{\rm J}$ in mass to have a radius of $\lesssim0.8\,R_{\rm J}$ given the H-He equation of state \citep{Chabrier1998, Chabrier2019, Chabrier2021, Chabrier2023}.

Intrinsically younger or lower mass objects will have lower surface gravities than the older field population of brown dwarfs \citep[e.g.][]{Faherty2016, Liu2016}, and since clouds become more prominent at lower surface gravity, these objects tend to have redder colors, cooler effective temperatures, and higher variability amplitudes \citep[e.g.][]{Liu2016, Vos2022}. The interplay between age, mass, and composition makes the atmospheric modeling of these objects particularly challenging. For example, the spectro-photometry of the HR~8799 planets \citep{Marois2008, Marois2010}, being much fainter and with much redder colors than their field spectral counterparts, have been challenging to model with physical consistency \citep[see][for updated GRAVITY spectroscopy of these planets and an exhaustive retrieval analysis]{Nasedkin2024}. At their discovery, the fidelity of the cloud model was seen to have an impact on the plausibility of the determined radii for these planets \citep[][supplimentary materials]{Marois2008}. Many early analyses of HR~8799 and other substellar objects found too small radii \citep[e.g.][]{Bowler2010, Barman2011a, Barman2011b, Marley2012}. For the young late L-type brown dwarf companion 2M1207b, this radius problem was rectified by increasing the complexity of the models: including cloud opacity with intermediate particle sizes and disequilibrium chemistry \citep{Barman2011b, Skemer2011}.  

For the remainder of this discussion, we make the distinction between ``physically implausible" radii and “inconsistent” radii. The former is a model estimated radius too small to realistically satisfy the substellar object's equation of state and thus the predictions from evolutionary models for any age. The latter is an atmospheric model estimated radius that is not consistent with the ones obtained with evolutionary tracks, but physically plausible. In this case discrepancies could arise from insufficient model physics in either class of model, or inaccuracies in the measurements of age, dynamical mass, etc for the planet. Since this section focuses on the atmospheric modelling for AF Lep b, we mostly discuss possible limitations of atmospheric models. We recognize however that missing information in evolutionary tracks might also contribute to ``inconsistent" radii. 

\subsection{Previous models of the planet's atmosphere} \label{subsec:discuss_prev_atmo}

In the case of AF~Lep~b, discovery studies found that it straddled the low-gravity L-T transition but individually did not have enough data to model the planet's atmosphere in much detail \citep{DeRosa2023, Franson2023c, Mesa2023}. Two subsequent studies collated the discovery observations and modeled this ensemble with \texttt{petitRADTRANS} retrievals \citep{Zhang2023}, and with the \texttt{Exo-REM} forward models \citep{Palma-Bifani2024}, and both encountered the small radius problem. \citet{Zhang2023}, by imposing a prior on the shape of the P-T profile based on the results of RCE forward models (their gradient P-T parameterization, described in \S\ref{subsec:retrieval_setup}), could effectively force the retrieved EddySed model to produce cloudy atmospheres. This is an elegant solution to the degeneracy between P-T flexibility and cloud opacity in retrievals, but did not appear to resolve the small radius problem (as our results also demonstrate). In an attempt to address the small radius problem, \citet{Zhang2023} imposed a prior on the radius based on evolutionary model predictions, restricting the allowed radii to values $R\in\mathcal{U}(1.2,1.5)$, as opposed to the uninformative prior commonly adopted when performing retrievals $R\in\mathcal{U}(0.5,2.0)$. \citet{Palma-Bifani2024} instead impose a prior on the surface gravity explicitly $\log(g)\sim\mathcal{N}(3.7,0.1)$, combining predictions from a single set of evolutionary models \citep{Marleau2019} and the dynamical mass estimate from \citet{Zhang2023}. The detection of the planet at $4.4\,\mu\mathrm{m}$ from \textit{JWST}/NIRCam \citep{Franson2024} enabled an updated retreival analysis using the same framework as in \citet{Zhang2023}. As their Figure 3 indicates, the relatively low $4.4\,\mu\mathrm{m}$ flux of the planet is due to out of equilibrium CO absorption; their retrievals find deep quenching pressures (running up against their prior boundary of 3, peaking near $2.80\,\mathrm{bar}$), high metallicities ($\mathrm{[Fe/H]}=1.67\pm0.20$), and super-solar C/O ($\mathrm{C/O}=0.65\pm0.05$). 

While AF~Lep~b lies in a temperature-gravity regime that is plagued by the small radii problem, both prior-based solutions appeared to resolve the issue. Armed with our new GRAVITY spectrum that sets a stronger temperature-pressure constraint due to the unambiguous methane absorption, we sought to revisit these results.

\subsection{Modeling the atmosphere of AF~Lep~b} \label{subsec:discuss_atmo}

\subsubsection{Forward model results}
\par We began our atmospheric analysis comparing forward models to only the GRAVITY spectrum (see Figure \ref{fig:gravonly_spectrum_fits}). The cloudless, chemical equilibrium \texttt{Sonora Bobcat} models provided a relatively poor fit to the data ($\chi^2_{\rm red}\sim4$, see Table \ref{tab:atmo_analysis}) and physically implausible radii. The cloudless, chemical disequilibrium \texttt{Sonora Elf Owl} models provided a better fit ($\chi^2_{\rm red}\sim2.2$) and indicate strong vertical mixing ($\log(K_{\rm zz})=7.7$) in broad agreement with the findings of \citet{Franson2024}, but still resulted in physically implausible radii. The cloudy, chemical disequlibrium \texttt{Exo-REM} models fit the GRAVITY spectrum well and resulted in a (multi-modal) posterior, where half of the posterior samples (in one mode) encompassed physically plausible radii and another half (in the other mode) were physically implausible. 

\par We then compared the entire SED of AF~Lep~b to the cloudy, solar metallicity, chemical equilibrium \texttt{BT-Settl} models, which provided a very poor fit to the data ($\chi^2_{\rm red}\sim6$). This was to be exepcted, as both the model fits to the GRAVITY spectrum alone as well as previous studies \citep{Zhang2023, Franson2024} indicate that an enhanced metallicity and chemical disequilibrium are required to explain the planet's spectrum. We then compared the full dataset to the \texttt{Exo-REM} models, first without restrictions on the radius/surface gravity, and then by effectively fixing the surface gravity to the expectations from evolutionary models, given our dynamical mass estimate. For the first case we find $T_{\rm eff}\sim850\,\mathrm{K}$, an enriched metallicity $\mathrm{[M/H]}=0.69\pm0.02$, and $\mathrm{C/O}=0.57\pm0.01$. The $\mathrm{log(g)}$ is higher than expected given the evolutionary models, and the mass is driven $\sim0.5\,M_{\rm J}$ higher than the prior, resulting in a radius that is physically plausible but inconsistent at $>5\,\sigma$ from the expectations from evolutionary models ($R_{\rm Exo-REM}=1.14\pm0.02\,R_{\rm J}$, compared to $R_{\rm SM08}=1.30\pm0.01\,R_{\rm J}$). This in turn leads to a higher effective temperature estimate than predicted by evolutionary models. In the second case, we fixed the mass to our dynamical mass central value and the surface gravity to $3.7$, which results in a $R=1.35\,R_{\rm J}$, provided the parallax is not driven away from its prior value. The parallax prior on the system from \textit{Gaia} is strong enough to hold, and we recover this physically plausible and consistent with evolutionary tracks radius in this case. Here we find a lower effective temperature in line with evolutionary model predictions $T_{\rm eff}\sim780\,\mathrm{K}$, a more enriched metallicity $\mathrm{[M/H]}=0.77\pm0.03$, and a slightly sub-solar $\mathrm{C/O}=0.49\pm0.01$. The difference in reduced goodness of fit between these models is $\Delta\chi^2_{\rm red}=-0.2$, slightly favoring the inconsistent model (as the analysis in \citet{Palma-Bifani2024} and Figure \ref{fig:fm_spectrum_fits} shows, it appears that this is primarily driven by the slope of the SPHERE spectrum and the magnitude of the Keck/NIRC2 \textit{L'} photometry). 

\par With \texttt{Exo-REM}, all the estimated radii are physically plausible, but it was not feasible to use our grid sampling scheme to recover consistent radii without fixing the surface gravity and mass a-priori (as in the second case). For instance, we attempted to place the prior $\log(g)\sim\mathcal{N}(3.7,0.1)$ as in \citet{Palma-Bifani2024}, which is less informative than simply fixing the value to 3.7 and more accurately captures our uncertainty on the system (and planet) age, inital conditions, etc., but the posterior distributions for these tests were always driven away from this prior range and to the equivalent of the posterior from the first case. Still, it is encouraging that the \texttt{Exo-REM} model comparisons resulted in derived radii that were larger than $1.0\,R_{\rm J}$. An improvement could be gained by varying the choice of the supersaturation parameter in the models, a ratio that, like $f_{\rm sed}$ in the \texttt{EddySed} model, informs the cloud thickness. In the models we used, the supersaturation parameter is fixed to an intermediate value of $S=0.03$, but its exact value shapes the \textit{Y}- and \textit{J}-bands and could reasonably vary anywhere between 0.001 and 0.1, based on the range of water nucleation in Earth's atmosphere \citep[see Figure 11 in][]{Charnay2018}. The public grid also does not include any sulfide salt clouds (like KCl) which may contribute additional wavelength dependent opacity to this region. Finally, the uncertainties we measure based on our method of grid comparison could benefit from the use of a spectral emulator, where instead of multi-linearly interpolating across spectra, the grid of pre-computed spectra is transformed via principle component analysis into eigenspectra, and the likelihood function can be modified to include the uncertainty due to the interpolation \citep{Czekala2015, Zhang2021a, Zhang2021b}. This results in more reasonable uncertainty estimates with wider posterior distributions on the sampled parameters, and might not require that we fix key parameters a-priori to recover physically consistent results.

\subsubsection{Retrieval results}

\par We constructed our retrieval experiments to incrementally increase the complexity of the cloud prescription (or P-T profile) and then add plausible physically motivated priors (like the RCE or radius priors from \citealt{Zhang2023}), in order to better understand the structure of the planets atmosphere and address or understand limitations of our models (primarily, the small radius problem). First, inspired by the suggestion in \citet{Tremblin2015}, we fit the data with a cloudless retrieval using the fully flexible cubic spline P-T profile and the (dis)equilibrium chemistry parameterization. This fit converged to a unphysical radius ($R=0.70\pm0.05\,R_{\rm J}$) with an isothermal photosphere. As in \citet{Zhang2023, Franson2024}, we retrieved a deep quenching pressure $P_{\rm quench}=2.67\pm0.06\,\mathrm{bar}$, indicating that vertical mixing was transporting hot, out of equilibrium gas from deep in the planet's atmosphere into the photosphere, but unlike those results, we find a more moderately enriched metallicity of $\mathrm{[Fe/H]}=0.83\pm0.08$. As we changed the retreival framework, the resulting small radius problem persisted, and in general the qualitative results remained the same: a moderately enriched metallicity, C/O near solar or slightly higher, a deep quench pressure.
\par We then fit the data with the single agnostic cloud model, the gradient P-T profile, and free chemistry again with a wide radius prior: while the P-T profile was influenced by the RCE priors, this also converged to an unphysical radius with an isothermal photosphere, which prompted us to add multiple cloud layers of varying specie using the EddySed model. We included 3 cloud species (Fe, MgSiO$_3$, and KCl), started with the flexible 3-part P-T profile, (dis)equilibrium chemistry, and a wide radius prior: this again converged to a physically implausible radius with an isothermal photosphere and relatively unconstrained cloud properties---notably, the quench pressure for this retrieval was slightly lower than the other values, but more uncertain, $P_{\rm quench}=2.28\pm0.22\,\mathrm{bar}$. 
\par We added the extra constraint of the gradient P-T profile paramterization with RCE based priors. These retrievals resulted in consistently cloudy atmospheres with more adiabatic P-T profiles, with a strong contribution from MgSiO$_3$ \citep[as in][]{Zhang2023, Franson2024}. We compared free vs disequilibrium chemistry using this gradient P-T parameterization, and while the fit to the (dis)equilibrium chemisty was marginally better ($\Delta\chi^2_{\rm red}=0.06$), the small radius problem persisted.

\par We moved on to using \citet{Zhang2023}'s evolutionary model informed uniform radius prior, $R\in\mathcal{U}(1.2,1.5)$, with the gradient P-T, (dis)eq. chemistry, and EddySed clouds: effectively this retrieval reproduced the setup in \citet{Zhang2023, Franson2024}, except that we include our new data and include GP parameters to estimate the correlation in the SPHERE spectrum. the posterior piled up against the lower bound of the radius prior range and the P-T structure of the photosphere was driven from the RCE priors towards a more isothermal shape, even when we placed a normally distributed prior on the radius based on the evolutionary models. This indicated to us that the EddySed model as currently implemented in \texttt{petitRADTRANS} is unequipped to explain our observations with physical consistency. 

\subsubsection{Small inferred radii and cloud particle sizes}
\par We constructed our retrieval experiments to incrementally increase the model cloud treatment's complexity and then add plausible, physically motivated priors, but the all retrievals we attempted yield physically implausible radii or posteriors that pile up onto the lower bound of the prior-space. Overall, the GRAVITY spectrum appears to set a stronger constraint on the correlated atmospheric parameters $T_\mathrm{eff}$ and $\log(g)$ than previously available $K$-band photometry. In this context, we can see that the restrictive radius prior from \citet{Zhang2023} was effectively a soft $T_\mathrm{eff}$, $\log(g)$ prior; previously available observations did not have sufficient statistical leverage to push the retrieval away from reproducing the prior. Our high quality GRAVITY data does, resulting in solutions that exhibit a physically implausible small radius again, even when a normally distributed prior is placed on the expected radius.

\par It could be that in the case of AF Lep~b, the root cause of the ``small radius problem” lies in the lack of diversity of cloud prescription across our models (see \S\ref{subsec:retrieval_setup}). We have a few tentative lines of evidence to support this. In \citet{Balmer2023} we found that with the EddySed cloud model within our \texttt{petitRADTRANS} retrievals we were unable to reproduce the expected solar metallicity of a massive brown dwarf companion (even when enforcing cloud opacity in the photosphere, albeit with a different prescription than the gradient P-T profile used here). In this case, the spectrum was well fit with the solar metallicity, chemical equilibrium \texttt{BT-Settl} grid that implements microphysics motivated cloud particle size distributions. As discussed in that work, the retrieval overcompensates for the cloud model's inability to capture both the shape of the \textit{Y}- and \textit{J}-bands (the wavelength dependent slope) and the overall opacity required to fit the NIR (the optical depth) simultaneously. Instead, the retrieval found some values for the EddySed cloud that reproduce the optical depth and then increased the opacity in the \textit{Y}- and \textit{J}-bands by raising the atmospheric metallicity. This indicates to us that the paramterization of the clouds may be too rigid in some aspects, but not flexible enough in other aspects to produce models with physical consistency. Additionally, the \texttt{Exo-REM} models do yield the best fits to our data with physically plausible (if not fully consistent) radii. This grid also implements microphysics motivated cloud particle size distributions, which  might be the reason underlying this better agreement. The particle sizes set by microphysical arguments typically have more small grains deeper in the atmosphere than the EddySed model, as we discuss below.

\par Recent papers have shown that the particle size distributions for the clouds in low gravity substellar objects diverges strongly from those given by the EddySed model prescription. This can be seen in Figure 2 from \citealt{Helling2008c}, or by comparing, for instance, Figure 6 in \citealt{Ackerman2001} and Figure 7 in \citealt{Samra2022}. More specifically, \citet{Samra2022} show that fragmentation becomes a dominant process in shaping the cloud particle size distribution in low surface gravity atmospheres, as collisions are driven by atmospheric turbulence. \citet{Luna2021} show that the mid-infrared silicate absorption feature observed in field brown dwarf spectra can be best fit using two populations of small grains, rather than the EddySed prescription where the size of the grain given pressure decreases monotonically. We might expect that the deep quench pressure (strong vertical mixing) implied by our GRAVITY spectrum and the \textit{JWST} photometry for AF~Lep~b contributes to enhanced particle-particle collisions within the atmosphere, especially given the planet's relatively low temperature and surface gravity. This would result in a run of mean particle size versus pressure that is much steeper (e.g. smaller particles throughout the photosphere down to the cloud deck) and varies non-linearly. The assumption of a log-normal particle size distribution could therefore be one of the limiting factors in our cloud parameterization \citep[in particular, see Figure 8 in][comparing particle size distributions from the CARMA framework, and see the discussion in the paragraph below regarding hazes, which might motivate more larger particles higher up in addition to the smaller particles throughout the photosphere]{Gao2021}.  Future work could look to incorporate the results of these theoretical studies into parametric retrievals, perhaps with interpolations across tables of pre-computed cloud properties derived from microphysical models, or with differently parameterized particle size distributions, or as priors on retrieved vertically varying parameters. 

\par Along these lines, it could be that the lower than expected radii indicate an inhomogeneous photosphere, for instance patchy clouds with varying degrees of coverage. This has been suggested as a potential improvement for retrievals suffering from the small radius problem \citep[e.g. \S6.2 in][]{Burningham2021}, but even in retrievals where a simple cloud coverage parameter is constrained and preferred over a purely homogeneous model, the small radius problem persists \citep{Vos2022}. 
\par Also, as noted by \citet{Burningham2021}, if the small radius indicates dark patches on an inhomogeneous atmosphere, they must be clouds of significantly different properties to those we retrieve based on our photospheric constraints, either having different particle size distributions and/or different compositions, perhaps very large fluffy particles high up to maintain a gray opacity that can suppress a significant fraction of the planet's flux. We note here that such an opacity source is rather suggestive of a high altitude haze, or more generally, of aggregate cloud particles in the upper atmospheric layers \citep[for a recent review, see][]{Vahidinia2024}. AF~Lep~b, being relatively close to its host compared to other direclty imaged planets, could experience a significant UV flux that could drive haze production, similar to those observed on Jupiter \citep[e.g.][]{Zhang2013}. 

\subsubsection{Adopted atmospheric parameters: strong evidence for moderate metallicities}

\par We see, as does other contemporary retreival work \citep[e.g.][]{Xuan2024b}, that the small radius problem does not necessarily invalidate the atmospheric abundances we derive, although their interpretation requires caution. Indeed, fixing the radius (when comparing the \texttt{Exo-REM} forward models) results in a difference in C/O of $0.08$ and [Fe/H] of $0.08\,\mathrm{dex}$ between the best fitting fixed radius or free radius case. The range of [Fe/H] across our 3 cloud, gradient P-T profile \texttt{petitRADTRANS} retrievals from the wide radius case to the restricted radius and normal prior radius case is negligible within the statistical uncertainties, and C/O varies by $0.1$ with statistical uncertainties of $0.05$. The similarity in the range of abundances between our forward model fits and retrievals also lends some credence to the values themselves. While it would be foolhardy to adopt one of the results and its statistical uncertainties as the ``true" value, we can reasonably adopt a range of values averaged over our model exploration and compare this to previous results and formation model expectations. 

\par Across our retrievals, we find in general an enriched metallicity ([Fe/H], $\mu_{1/2}\in[0.3-0.9]$, $\mu_{1\,\sigma}\sim\pm0.07]$) 
that is in agreement with \citet{Palma-Bifani2024} and our \texttt{Exo-REM} forward model best fits. Our revised metallicity disagrees with the very enriched values found by previous \texttt{petitRADTRANS} retrievals \citep[$\mathrm{[Fe/H]}{=}1.67^{+0.17}_{-0.21}$][]{Zhang2023, Franson2024}. These previous results were obtained with a low resolution spectroscopy up to 1.6 micron and photometry redwards, in \textit{K}-band, \textit{L'}-band, and then F444W. Clouds were enforced using the prior restricted gradient P-T parametrization and physically motivated uniform radius priors (similar to last line of Table 4). This very enriched value appeared to be in tension at ${\sim}1\,\sigma$ with the planetary mass metallicity relationship \citep[][and see \S\ref{subsec:discuss_formation}, Figure \ref{fig:mass-metallicity}]{Thorngren2016}. By directly constraining the methane abundance with the GRAVITY spectrum and more strongly constraining the cloud wavelength slope with the updated SPHERE spectrum, our observations have elucidated this tension. One way the retrievals can suppress the flux of the atmosphere to match the data is by turning up the metallicity. Without strong constraints on the cloud properties, nor a high enough signal methane constraint, but with a restriction to consistent radii (as discussed above), the previous retrievals appear to have accepted solutions where the metallicity is exceptionally enhanced. In this paper, regardless of the radius prior, our data sets a strong enough constraint to systematically revise the metallicity towards more moderately enriched values (but still statistically significantly enriched compare to the stellar value, see \S\ref{subsec:discuss_formation}). Our revised metallicity agrees exceptionally well with the planetary mass metallicity trend.

\par We find C/O ratios that range about the solar value ($\mu_{1/2}\in[0.43-0.61]$, $\mu_{1\,\sigma}\sim\pm0.05$). We find effective temperatures ($T_{\rm eff}$, $\mu_{1/2}\in[780-980\,\mathrm{K}]$, $\mu_{1\,\sigma}\sim\pm20\,\mathrm{K}$) that bracket previous estimates, depending on model and radius prior choice. In our retrievals where we parameterize the effects of disequilibrium chemistry using a quench pressure, we find large (deep) quench pressure values ($P_{\rm quench}$, $\mu_{1/2}\sim2.4-2.8\,\mathrm{bar}$, $\mu_{1\,\sigma}\sim\pm0.05\,\mathrm{bar}$), indicating uniform and out of equilibrium abundances of H$_2$O, CO, and CH$_4$ in the photosphere. This is in good agreement with \citet{Zhang2023} and \citet{Franson2024}, although our revised values are slightly lower than theirs, due to our lower metallicity estimate. The retrievals also indicate that because of the degeneracy with the restrictive radius prior and P-T profile, the exact estimation of $P_{\rm quench}$ is influenced by the small radius problem. Hopefully, higher resolution measurements with HiRISE \citep[e.g.][]{Vigan2023} or KPIC \citep[e.g.][]{Xuan2024b} will be able to probe a wider range of pressures and their abundance estimates will therefore be less sensitive to these degeneracies. Spectral observations at longer wavelengths with \textit{JWST} (e.g. GO-5342, PI: Xuan) will help to better constrain the abundances of C and O bearing molecules. Given the improvement in the radius determination between \texttt{Exo-REM} and other forward models, it is reasonable to expect that changes in the cloud modeling (considering particle size distributions better informed by microphysics) or inhomogenous clouds in retreivals could help alleviate the small radius problem and enable better constraints on the planet's fundamental properties.

\par For now we adopt, from a combination of the median of the range our forward models and retrievals and inflating the typical uncertainties on the median by a factor of two, the following parameter range estimates for the atmopshere of AF~Lep~b: $T_{\rm eff}=800\pm50$, $\log(g)=3.7\pm0.2$, $R=1.3\pm0.15\,R_{\rm J}$, $\mathrm{[Fe/H]}=0.75\pm0.25$, $\mathrm{C/O}=0.55\pm0.10$, $P_{\rm quench}=2.6^{+0.2}_{-0.3}$. 

\subsection{Hints at formation history from an initial atmospheric analysis} \label{subsec:discuss_formation}

\par Formation via core accretion (or within a disk more generally) could encode a specific atmospheric C/O and metallicity, depending on the location of formation in relation to various snowlines, where gaseous molecules condense into solid icy grains \citep{Oberg2011}, and the subsequent accretion history of the planet. The actual measurement of these quantities, especially C/O, via atmospheric modeling has proven difficult \citep[see discussion in][]{Hoch2023}, even for benchmark brown dwarf companions with better constrained fundamental properties \citep[see e.g.][]{Line2015, Wilcomb2020, Calamari2022, Wang2022, Xuan2022, Rowland2023, Balmer2023, Xuan2024b, Calamari2024}. Nevertheless, significant progress has been made in the past decade on directly imaged objects \citep[e.g.][]{Konopacky2013, Molliere2020, jaWang2020, Wang2021b, Wang2023, Hoch2023, Nasedkin2024}. Provided the measurement can be made accurately enough for the planet, the interpretation of planetary C/O ratios is especially sensitive to planetary formation model assumptions \citep{Molliere2022}, as well as to measurements of the baseline host star abundances \citep{Reggiani2022, Reggiani2024}. 

\par
\begin{figure}
    \centering
    \includegraphics[width=0.75\textwidth]{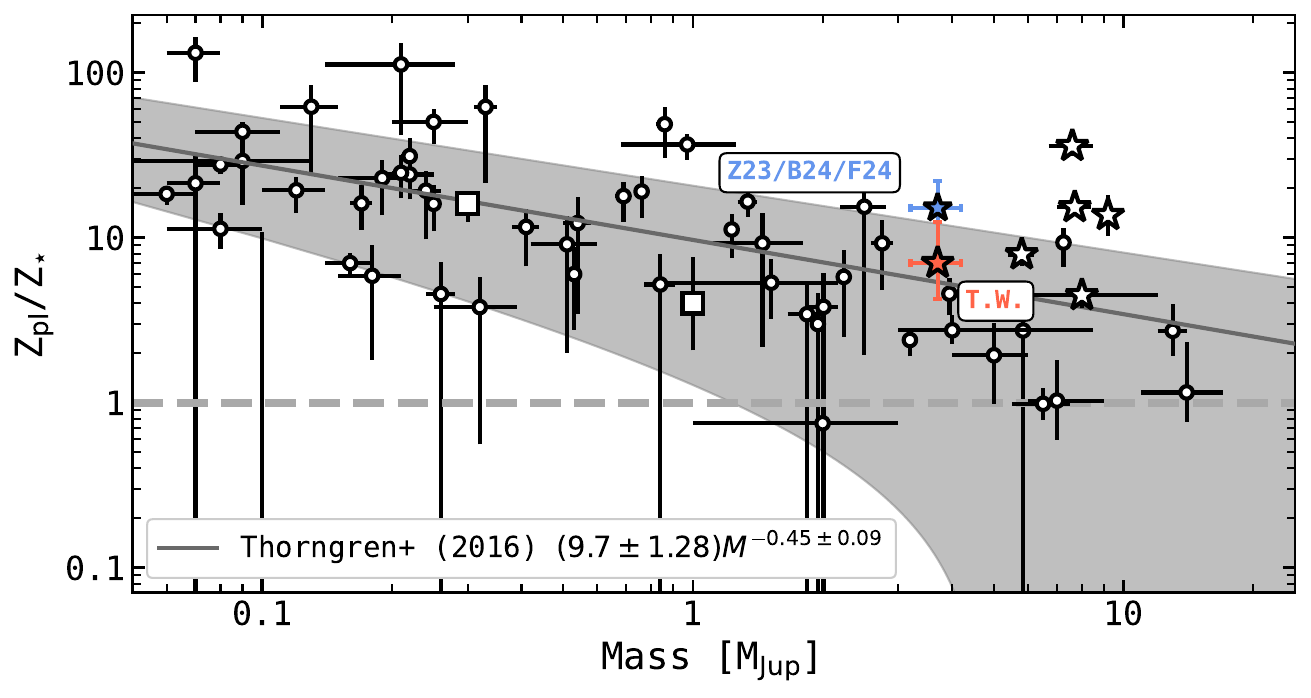}
    \caption{The mass-metallicity relationship for giant exoplanets, adapted from \citet{Thorngren2016} and \citet{Nasedkin2024}, references therein. AF~Lep~b from this work is indicated by a red star while AF~Lep~b with the mass from \citet{Bonse2024} and metallicity from \citet{Zhang2023, Franson2024} is indicated by a blue star. Other planets that have metallicities estimated based partially on GRAVITY spectra and \texttt{petitRADTRANS} retrievals, namely $\beta$-Pictoris~b \citep{GravityCollaboration2020} and the HR~8799 planets \citep{Molliere2020, Nasedkin2024}, are indicated with black stars. The Solar System gas giants, Jupiter and Saturn, are indicated with squares, with bulk abundances from \citet{Guillot1999, Thorngren2016}. Note that, unlike the other error bars that only include the uncertainty on the planetary abundances, the uncertainties for AF~Lep~b include the uncertainty on the stellar abundances and the planetary abundances combined.}
    \label{fig:mass-metallicity}
\end{figure}

 The planet's metallicity ($\mathrm{[M/H]}=0.75\pm0.25$) appears straightforward to interpret in the context of core accretion. Transiting giant exoplanets appear to follow a mass-metallicity relationship \citep{Guillot2006, Miller2011, Thorngren2016} consistent with expectations from core accretion \citep{Hasegawa2018}. In order to transform our planetary [M/H] estimates into a ratio of $Z_\mathrm{planet}/Z_\star$ we assume the stellar metallicity follows the $\beta$~Pic moving group proxy metallicity ([Fe/H]$=+0.13\pm0.10$) derived in \citet{Reggiani2024}. This estimate agrees with the stellar photospheric metallicity of AF~Lep~A ([Fe/H]$=-0.27\pm0.31$) derived in \citet{Zhang2023} at $2\,\sigma$, but we use the proxy metallicity here because the photopsheric abundances of rapidly rotating early type stars are more difficult to estimate than their later type siblings, and the dispersion of metallicities in open clusters is generally of order $0.03$ (\citealp{Poovelil2020}, and see the discussion in \S6 of \citealp{Reggiani2024}). Following \citet[][equations 2-3]{Thorngren2019} and \citet[][equations 18-20]{Nasedkin2024}, and propagating the uncertainty on the host star metallicity, we transformed $\mathrm{[M/H]}=0.75\pm0.25$ into $Z_{\rm pl}=0.073^{+0.050}_{-0.031}$ assuming the planet is fully mixed (i.e. has no solid core) and derived $Z_\mathrm{pl}/Z_\star=7.0^{+5.5}_{-2.8}$. With the dynamical mass from our orbit fit, we plotted AF~Lep~b in a mass-metallicity diagram with other exoplanets and directly imaged companions (Figure \ref{fig:mass-metallicity}). For massive planets ($\gtrsim2\,M_{\rm J}$) the assumption that they are fully mixed, so that their atmospheric metallicity corresponds to their bulk metallicity, is not unfounded \citep[but see discussion in ][]{Thorngren2019}. AF~Lep~b shows excellent agreement with the empirical relation fit to warm transiting Juptiers in \citet{Thorngren2016}.

\par The adopted range of C/O encompasses the olar value, and the adopted metallicity indicates the planet is metal rich. Given these facts, we make a highly simplified, illustrative argument about the formation history of AF~Lep~b. We used the formation inversion model presented in \citet{Molliere2022} assuming the static protoplanetary disk model from \citet{Oberg2011}. This model indicates that the planet likely accreted its solids beyond the CO iceline, but sets no constraint on where the gas was accreted. In this case, the planetary core might have formed beyond the CO iceline and migrated inwards. It is important not to over-interpret this result given the simplistic disk model, however. In \citet{Molliere2022} it was shown that in the case of HR~8799~e, with a similar abundance constraint (moderately enhanced [Fe/H], solar C/O), including the effect of chemical evolution within the disk model resulted in an entirely different formation interpretation; the planet's current orbit and abundance pattern was consistent with in-situ formation. Stronger constraints on the planet's abundances or constraints on isotopologues \citep[e.g.][]{Xuan2024b} could motivate more detailed formation inversion modeling. Future work should treat the planet formation inversion modeling with more detailed protoplanetary disk models (for instance, models that vary abundances over time as the gas evaporates, or as large dust grains drift inwards), and employ a more careful reading of the planetary to stellar abundance ratio as the planetary C/O becomes more refined (for instance, \citealt{Reggiani2024} find that their $\beta$~Pic abundance proxy HD~181327 has a super-solar C/O$=0.62\pm0.08$, which encompasses our estimates at $1-3\,\sigma$, depending on the model).

\subsection{Constraints on the properties of planets observed with only absolute astrometry and GRAVITY} \label{subsec:discuss_grav+gaia}

\par As in \citet{Winterhalder2024}, we find that coupling absolute astrometry with relative astrometry from GRAVITY results in well constrained posterior distributions of orbital elements. An orbit fit to the HGCA and our 3 epochs of GRAVITY relative astrometry (taken over only two months) results in a posterior distribution effectively equivalent to that recorded in Table \ref{tab:orbit} but with wider uncertainties: semi-major axis constrained to within $\pm0.5\,\mathrm{au}$, a $1\,\sigma$ upper limit on the eccentricity of 0.05, and an inclination angle to within $2^\circ$. This indicates that \textit{Gaia} discovered planets directly observed uniquely with GRAVITY will have well constrained orbital elements \citep[see also the BD examples in][]{Winterhalder2024}. Our (dis)equilibrium chemistry, agnostic cloud, gradient P-T retrieval with a wide radius prior, fit to only the GRAVITY spectrum, found an unphysically small radius $R=0.70\pm0.05\,R_{\rm J}$, slightly sub solar C/O$=0.50\pm0.05$, and a deep quench pressure $P_{\rm quench}\sim2.67\pm0.06\,\mathrm{bar}$,  consistent within $1-2\,\sigma$ with the equivalent retrievals performed on the entire SED. Our free chemistry, agnostic cloud, gradient P-T retrieval with a wide radius prior, fit to only the GRAVITY spectrum, yielded constraints on the log mass fraction of H$_2$O$=-2.86\pm0.05$, CO$=-2.99\pm0.10$, and CH$_4=-4.99\pm0.08$, that is, constraints on the individual molecular abundances to about 5\% statistical uncertainty. We find a recombined C/O$=0.43\pm0.06$ for this retrieval, the lowest C/O in our retrieval analysis, which could indicate some missing oxygen contribution from, for instance, solid grains (but this could be solved by a retrieval with greater cloud model fidelity). The cloud properties are unconstrained by the $K$-band spectrum and result in a largely cloudless atmosphere in both cases, but interestingly, the free chemistry retrieval resulted in a radius ($R=1.22_{-0.14}^{+0.19}\,R_{\rm J}$) that is consistent within $1\,\sigma$ with the expectations from evolutionary models. This indicates that the parameterization of (dis)equilibrium chemistry is more deterministic of the bulk properties of the atmosphere than the free chemistry parameterization. The cloudy \texttt{Exo-REM} forward models fit to only the GRAVITY spectrum resulted in multible solutions, some of which had physically plausible radii. Regardless of current atmospheric model shortcomings, these are very encouraging results for exoplanet science coupling \textit{Gaia} and GRAVITY, as future observations should be high enough quality to facilitate comparative studies leveraging the orbital elements \citep[e.g. Figure 17,][]{Bowler2020} or molecular abundances (provided the systematic uncertainty in the atmospheric results can be marginalized over when comparing between planets fit using the same model framework).

\section{Conclusions} \label{sec:conclusion}

\par We obtained new spectroscopic \textit{K}-band ($R=500$, $1.9-2.5\,\mu\mathrm{m}$) interferometric observations of the directly imaged planet AF~Lep~b using VLTI/GRAVITY, which achieved $\sim50\,\mu\mathrm{as}$ relative astrometric precision and resulted in the direct detection of methane absorption in the planet's atmosphere. We also re-reduced the SPHERE IFS ($R=30$, $0.95-1.64\,\mu\mathrm{m}$) spectrum of the planet from \citet{Mesa2023} using the \texttt{robust PACO} algorithm. 
\par These observations allowed us to measure the orbit of the planet to high precision. The planet's revised dynamical mass is $M_{\rm b}={3.75}_{-0.5}^{+0.5}\,M_{\rm J}$, making AF~Lep~b one of the lowest mass directly imaged planets, and consistent with the $<4\,M_{\rm J}$ population of radial-velocity planets orbiting metal rich host stars \citep[a property associated with core accretion formation, ][]{Santos2017, Schlaufman2018}. We determined that the planet's orbit is circular to within an upper limit of $e<0.02, 0.07, 0.13$ at $1, 2, 3\,\sigma$ confidence, and that the planet's orbital inclination ($i_{\rm orbit}={57.5}_{-0.7}^{{+0.6}^\circ}$) is consistent with being aligned with the host star's rotation axis $i_{\star}=54_{-9}^{{+11}^\circ}$. A circular, aligned orbit is the expected outcome of formation within a protoplanetary disk, and there appears to be growing population-level evidence that supports the hypothesis that directly imaged planets follow this expected outcome \citep{Bowler2020, Bowler2023, Sepulveda2024}. In light of a non-detection of additional, more widely separated gas giant planets from \textit{JWST} \citep{Franson2024}, we show that given our orbit constraints, ${<}1/3$ of accepted orbits are sufficiently eccentric to ``stir" the outer debris belt at $46\pm9\,\mathrm{au}$; the maximum a-posteriori orbit is inconsistent with stirring the debris, and would necessitate another disk driven process, or additional smaller planets to explain the debris population.

\par New \textit{K}-band spectroscopy of the planet facilitated an updated atmospheric analysis that represents a departure from previous studies in a few key ways. Reproducing the forward model analysis in \citet{Palma-Bifani2024} and the retrieval analysis in \citet{Zhang2023} and adding the new GRAVITY spectrum results in derived radii that are inconsistent with expectations based on the dynamical mass and evolutionary models (and in some cases, physically impossible radii); it was necessary to fix the radius a-priori to recover physically plausible solutions. Nevertheless, noting the general agreement of our atmospheric abundances regardless of the modeled radii, marginalizing over the radius problem (and over the differences between the forward models and retrievals), we determined an updated range of atmospheric parameters for the planet: $T_{\rm eff}=800\pm50$, $\log(g)=3.7\pm0.2$, $R=1.3\pm0.15\,R_{\rm J}$, $\mathrm{[Fe/H]}=0.75\pm0.25$, $\mathrm{C/O}=0.55\pm0.10$. In the context of the planetary mass-metallicity trend AF~Lep~b coincides with the relationship fit to the bulk metallicities of warm transiting Jupiters identified in \citet{Thorngren2016}, a trend which is associated with core accretion formation \citep{Hasegawa2018}. We touch briefly on potential future work connecting the atmospheric abundances of the planet with formation models, but we caution that the major limiting factors here are still systematic uncertainties in our atmospheric models and limiting, simplifying assumptions in the protoplanetary disk models that can be considered in this inversion framework \citep[see discussion in][]{Molliere2022}. 

\par This work also demonstrates that future observations of ice-line separation planets, detected with \textit{Gaia} absolute astrometry in DR4 and uniquely directly imaged with VLIT/GRAVITY, can constrain the orbits of these planets and the atmospheric abundance of key molecular species. 
In the near future, continued astrometric monitoring of this planetary system could reveal higher order signals, like epicycles induced by inner giant planets \citep{Lacour2021} or particularly massive exomoons, while continuing to refine the planet's eccentricity upper limit to assess the stirring of the debris disk. In the next decade, long baseline optical interferometry will enable a census of the formation histories of young gas giant planets detected with absolute astrometry, and will better inform our understanding of where, when, and how gas giant planets form. 

\begin{acknowledgements}
\section*{Acknowledgements}
\par Many thanks to our night astronomers for these service mode observations, Claudia Paladini, Abigail Frost, Julien Drevon, and Thomas Rivinius, and to the telescope operators Alex Correa, Leonel Rivas, and Rodrigo Palominos. We are infinitely grateful to the Paranal and ESO staff for their support. Thanks also to Tim Pearce for discussion of planet-debris interactions. We thank the anonymous reviewer for their constructive response, which helped us improve this manuscript.
\par This work is based on observations collected at the European Southern Observatory under ESO programmes 112.25GE.001, 112.25GE.002, 112.25GE.003.
\par Part of this work was carried out by W.O.B. at the Advanced Research Computing at Hopkins (ARCH) core facility (rockfish.jhu.edu), which is supported by the National Science Foundation (NSF) grant number OAC1920103.
\par This work used the Dutch national e-infrastructure with the support of the SURF Cooperative using grant no. EINF-1620.
\par K.F.\ acknowledges support from the National Science Foundation Graduate Research Fellowship Program under Grant No. DGE 2137420.  T.S.\ acknowledges the support from the Netherlands Organisation for Scientific Research (NWO) through grant VI.Veni.202.230.
\par S.L.\ acknowledges the support of the French Agence Nationale de la Recherche (ANR), under grant ANR-21-CE31-0017 (project ExoVLTI). Z.Z.\ is supported by NASA Hubble Fellowship grant HST-HF2-51522.001-A. J.J.W., A.C., and S.B.\ are supported by NASA XRP Grant 80NSSC23K0280.
\par This research has made use of the VizieR catalogue access tool, CDS, Strasbourg, France (DOI: 10.26093/cds/vizier).
\par This research has made use of the Jean-Marie Mariotti Center \texttt{Aspro} service.
\par This publication makes use of data products from the Two Micron All Sky Survey, which is a joint project of the University of Massachusetts and the Infrared Processing and Analysis Center/California Institute of Technology, funded by the National Aeronautics and Space Administration and the National Science Foundation. 
\par This work has made use of data from the European Space Agency (ESA) mission {\it Gaia} (\url{https://www.cosmos.esa.int/gaia}), processed by the {\it Gaia} Data Processing and Analysis Consortium (DPAC, \url{https://www.cosmos.esa.int/web/gaia/dpac/consortium}). Funding for the DPAC has been provided by national institutions, in particular the institutions participating in the {\it Gaia} Multilateral Agreement.
\par This project has received funding from the European Research Council (ERC) under the European Union's Horizon 2020 research and innovation programme (COBREX; grant agreement n$^\circ$ 885593). 
\par SPHERE is an instrument designed and built by a consortium consisting of IPAG (Grenoble, France), MPIA (Heidelberg, Germany), LAM (Marseille, France), LESIA (Paris, France), Laboratoire Lagrange (Nice, France), INAF - Osservatorio di Padova (Italy), Observatoire de Genève (Switzerland), ETH Zürich (Switzerland), NOVA (Netherlands), ONERA (France) and ASTRON (Netherlands) in collaboration with ESO. SPHERE was funded by ESO, with additional contributions from CNRS (France), MPIA (Germany), INAF (Italy), FINES (Switzerland) and NOVA (Netherlands). SPHERE also received funding from the European Commission Sixth and Seventh Framework Programmes as part of the Optical Infrared Coordination Network for Astronomy (OPTICON) under grant number RII3-Ct-2004-001566 for FP6 (2004-2008), grant number 226604 for FP7 (2009-2012) and grant number 312430 for FP7 (2013-2016). 
\par This work has made use of the SPHERE Data
Centre, jointly operated by OSUG/IPAG (Grenoble), PYTHEAS/LAM/CeSAM (Marseille), OCA/Lagrange (Nice), Observatoire de Paris/LESIA (Paris), and Observatoire de Lyon (OSUL/CRAL).

\par WOB acknowledges their cat, Morgoth, for her ``encouragement."

\end{acknowledgements}

\appendix



\section{Additional posterior distributions}

\par
\begin{figure}
    \centering
    \includegraphics[width=\textwidth]{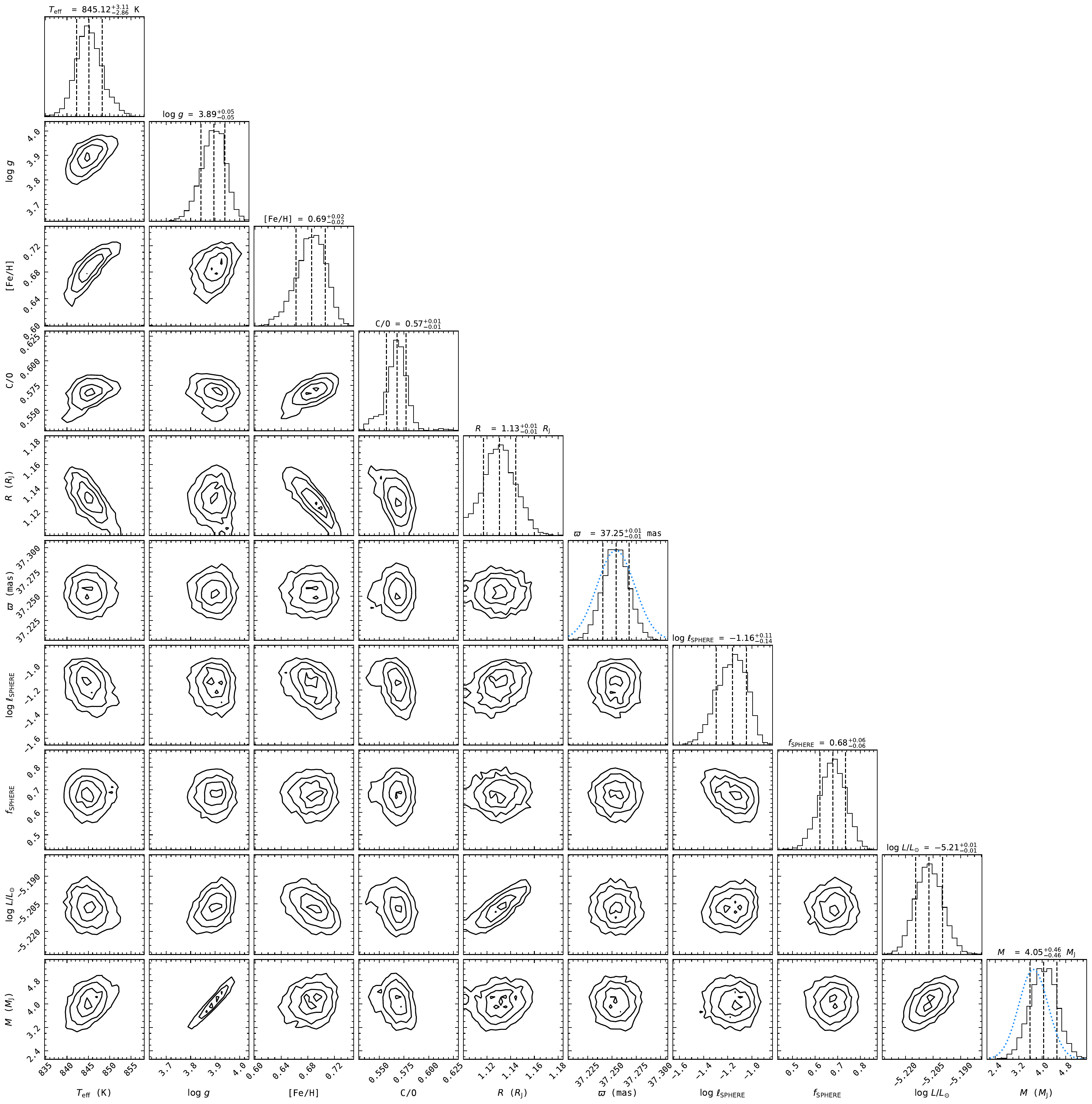}
    \caption{The posterior distribution of atmospheric parameters for the \texttt{Exo-REM} model fit to our observations for the case where $\log(g)$ is not fixed to 3.7, but we adopt the dynamical mass as a prior and let $\log(g)$, $R$ vary. Light blue dashed normal distributions plotted over the 1-D marginalized posteriors for parallax ($\varpi$) and planetary mass ($M$) denote the adopted priors as explained in \S\ref{sec:spec_analysis}.}
    \label{fig:exo-rem_post}
\end{figure}

\begin{figure}
    \centering
    \includegraphics[width=\textwidth]{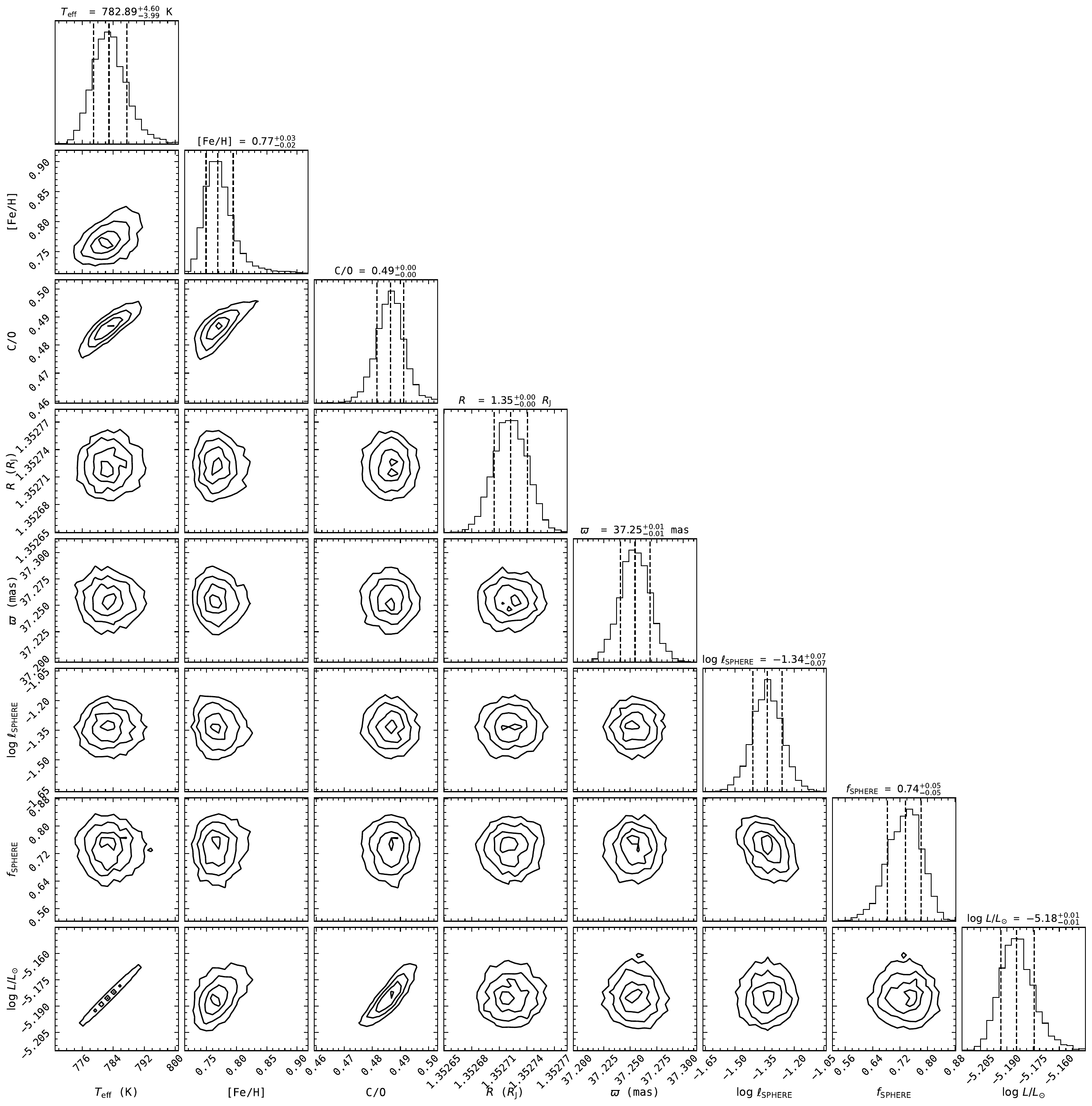}
    \caption{The posterior distribution of atmospheric parameters for the \texttt{Exo-REM} model fit to our observations for the case where $\log(g)$ is fixed to 3.7, the mass is fixed to the central value of our dynamical mass, and teh resulting radius is $1.35\,R_{\rm J}$. Light blue dashed normal distributions plotted over the 1-D marginalized posteriors for parallax ($\varpi$) as explained in \S\ref{sec:spec_analysis}. Unlike Figure \ref{fig:exo-rem_post}, $\log(g)$ and mass are not shown, as they were not varied.}
    \label{fig:exo-rem_fixedlogg_post}
\end{figure}

 This appendix includes plots representing the posterior distributions for the forward models we present in this paper. The posterior distribution for our \texttt{Exo-REM} atmospheric model fit to all observations using \texttt{species} is shown in Figure \ref{fig:exo-rem_post}, and the fit where we fix the surface gravity to 3.7 is shown in Figure \ref{fig:exo-rem_fixedlogg_post}.

\newpage 
\bibliography{aflepb_gravity}{}
\bibliographystyle{yahapj}  



\end{document}